\documentclass[a4paper]{article}

\usepackage{a4wide}
\usepackage[USenglish]{babel}
\usepackage{graphicx,subfigure}
    \graphicspath{{./}{figure/}{figure/indep/}{figure/dep/}{figure/pop/}}
\usepackage[colorlinks=true,linkcolor=blue,citecolor=red]{hyperref}
\usepackage{amsmath,amsthm,bbold}
\usepackage[inline]{enumitem}

\theoremstyle{remark}\newtheorem{remark}{Remark}[section]

\DeclareMathOperator*{\esssup}{ess\,sup}
\newcommand{\abs}[1]{\left\lvert#1\right\rvert}
\newcommand{\ave}[1]{\left\langle#1\right\rangle}
\newcommand{\cB}{\mathcal{B}}
\newcommand{\Beta}{\operatorname{B}}
\newcommand{\cC}{\mathcal{C}}
\newcommand{\cD}{\mathcal{D}}
\newcommand{\cE}{\mathcal{E}}
\newcommand{\norm}[1]{\left\Vert#1\right\Vert_\infty}
\newcommand{\op}{\mathrm{op}}
\newcommand{\cP}{\mathcal{P}}
\newcommand{\pop}{\mathrm{pop}}
\newcommand{\pr}[1]{{}^\prime\!#1}
\newcommand{\R}{\mathbb{R}}

\allowdisplaybreaks

\begin{document}
\title{Opinion modeling on social media and marketing aspects}

\author{Giuseppe Toscani\thanks{Department of Mathematics ``F. Casorati'', University of Pavia, and IMATI CNR, Via Ferrata 1, 27100 Pavia, Italy
            (\texttt{giuseppe.toscani@unipv.it})} \and
        Andrea Tosin\thanks{Department of Mathematical Sciences ``G. L. Lagrange'', Politecnico di Torino, Corso Duca degli Abruzzi 24, 10129 Torino, Italy
            (\texttt{andrea.tosin@polito.it})} \and
        Mattia Zanella\thanks{Department of Mathematical Sciences ``G. L. Lagrange'', Politecnico di Torino, Corso Duca degli Abruzzi 24, 10129 Torino, Italy
            (\texttt{mattia.zanella@polito.it})}}
\date{}

\maketitle

\begin{abstract}
We introduce and discuss kinetic models of opinion formation on social networks in which the distribution function depends on both the opinion and the connectivity of the agents. The opinion formation model is subsequently coupled with a kinetic model describing the spreading of popularity of a product on the web through a social network. Numerical experiments on the underlying kinetic models show a good qualitative agreement with some measured trends of hashtags on social media websites and illustrate how companies can take advantage of the network structure to obtain at best the advertisement of their products.
\end{abstract}


\section{Introduction}
Social media are nowadays an important  transmission vehicle of information. In the last twenty years we are witnessing an explosion of Internet-based messages transmitted through these media. They have become a major factor in influencing various aspects of consumer behavior including awareness, information acquisition, opinions, attitudes, purchase behavior, to name a few~\cite{Kaplan,Mangold,pan2010NJP,redner1998EPJB,sinha2005CHAPTER,sinha2004EPJB}. Therefore, more and more companies started to make use of social media in promotional efforts. On the other hand, maybe in reason of the recent development of these commercial strategies, even though social media is magnifying the impact that consumer-to-consumer conversations have in the marketplace, consolidated methods for increasing the impact of products through social networks have not yet been articulated.

Kinetic models of opinion formation in a multi-agent system of individuals characterized also by further parameters have been addressed in various papers. Among others, the modeling of the first part of this paper has points of contact with a recent work by D\"uring et al.~\cite{duering2015}, cf. also~\cite{Duering2009}. There, the opinion variable is coupled with a further parameter denoting the assertiveness of the agents (with high assertiveness corresponding to leadership). Also the dynamics of opinion formation in a society with a marked presence of zealot or stubborn individuals, i.e., agents who tend to maintain their strong opinions after interacting with other agents, have been dealt with in a number of papers. The effect of conviction was studied mainly in discrete models of opinion dynamics related to consensus formation, voting dynamics, game theory models, and diffusion of innovations among other applications~\cite{Biswas2011,brugnaPRE,Crokidakis2013,lipowski,Mobilia2007,waagenPRE,xiePRE}. In these works it is shown, mainly through simulations, how agents with a certain stubbornness can affect the process of consensus formation, especially as far as the kind of expected equilibria that could arise due to their influence and the time needed to convergence are concerned. Last, to stress the importance of understanding opinion dynamics in the modern societies, we quote the recent attempts to investigate the process of opinion formation in presence of uncertain interactions between agents~\cite{tosin2018CMS} and to act on it by means of suitable control strategies~\cite{Albi2014,Albi2016a}.

In recent years, the empirical study of social networks and of their role in the formation and spreading of opinions gained a lot of momentum thanks to the extraordinary large amount of data coming from online platforms. In particular, large-scale aggregate statistical descriptions showed patterns and periodic structures in the connectivity of real world networks~\cite{clauset2009SIREV,Liu2016,newman2002PNAS}. In this article, resorting to the powerful methodology of statistical physics~\cite{Albert2002,pareschi2013BOOK}, we will try to explore some of the aforementioned issues about the popularity dynamics on social media by introducing mathematical models of kinetic type able to follow the marketing of products on social networks. Recent researches on marketing aspects of social media~\cite{Mangold} revealed that social media are considered an important component of the promotional strategy, and therefore are incorporated as an integral part of the marketing strategy. Mangold and Faulds~\cite{Mangold} describe social media as a \textit{hybrid element of promotion mix} for its dual marketing functions. First, social media can be utilized as a traditional integrated marketing communications tool (e.g., direct marketing), where companies control the content, timing, and frequency of information being shared with consumers. Secondly, social media enables consumers to communicate with each other within their social networks, which creates a further attracting effect for companies. This hybrid marketing tool brings a new challenge to marketers, because they need to learn how to effectively spread information on a product over the largest audience of their target consumers. While discussing about possible strategies, Kaplan and Haenlein~\cite{Kaplan} suggested to deeply investigate social processes to better understand the social networks behavior. Accordingly, it seems appropriate in a first step to study opinion formation on social networks. 

Unlike classical modeling of opinion formation in a multi-agent society~\cite{Aletti2007,Ben-Naim2005,Bordogna2007,Boudin2009,cristiani2018MMS,toscani2006CMS}, where agents are considered indistinguishable, here we will take into account a further parameter, the network range, or \textit{connectivity}, of the agents, which can be reasonably measured in terms of followers. Highly connected agents are identified as \textit{influencers}, because, thanks to their large number of followers, their opinions can reach and influence many other users of the social media~\cite{Freberg2011}. As a matter of fact, however, influencers obey the same mechanism of opinion sharing as any other agent. In particular, they are not necessarily recognized as leaders and typically do not operate in a coordinated manner. The key point is simply the assumption that the opinion of the agents with a high connectivity, i.e., a large number of followers, results more convincing than that of the agents with a low number of followers. As a main difference with~\cite{albi2017KRM,Barre2018}, where the authors consider dynamic networks, here we structure the population of the agents by means of a fixed background network with realistic connectivity distribution.

Starting from these assumptions, in Section~\ref{sect:opinion.dynamics} we will introduce a kinetic equation modeling opinion formation in presence of the connectivity parameter. Then, in Section~\ref{sect:popularity} we will couple to such kinetic model a further kinetic equation describing the time evolution of the popularity of a certain product. Finally, in Section~\ref{sect:numerics} we will perform numerical experiments based on the previous model equations, trying to ascertain the better strategy that a company can find to promote at best its products.

The analysis performed in the present work enlightens that the role of an underlying social network is of paramount importance in the formation of opinion patterns among the agents, such as e.g., consensus/dissent. Moreover, it clearly demonstrates that highly connected agents are fundamental to increase in time the popularity of a product, thereby allowing it to be sold at best. The mechanism described by the coupling of the opinion formation model with the popularity spreading model is such that it gives, in some cases, apparently counterintuitive but finally realistic answers. Indeed, it can happen that the natural choice to spread out the message on a large number of agents with low connectivity, while giving an immediate increase in popularity, does not actually guarantee the persistence of the popularity in time. On the other hand, by addressing few people with high connectivity the initial decay of the popularity turns out to be only a local effect that connectivity will subsequently remedy.

\section{Kinetic modeling of opinion dynamics over networks}
\label{sect:opinion.dynamics}
\subsection{Microscopic binary model}
In the socio-physics community, a customary procedure for modeling the formation of opinions in a population of agents consists in representing the opinion of an individual, with respect to a certain subject, by a real number. This number can vary in some discrete set or in a fixed interval, say $[-1,\,1]$, where $\pm 1$ denote the extremal opinions. Individual changes of opinion are assumed to be a result of random binary interactions between pairs of agents. Specifically, the pre-interaction opinions $w$ and $w_\ast$ of two agents will turn into the new post-interaction opinions $w'$ and $w_\ast'$ as a consequence of the discussion opened by the two agents, of the influence of external factors such as media or propaganda, and of the spontaneous self-thinking~\cite{pareschi2013BOOK,toscani2006CMS}. 

As briefly discussed in the introduction, modeling opinion dynamics over a social network naturally requires to couple the opinion variable with further parameters able to characterize the \textit{range} of the agents in the social network. Consequently, we assume that the microscopic state of each agent is given by the pair $(w,\,c)$, where $w\in [-1,\,1]$ is the opinion of the agent and $c\in\R_+$ is his/her \textit{connectivity} in the social network, represented e.g., by the number of followers. 

The connectivity parameter $c$ is a measure of the credibility of an agent. The higher $c$, the higher the credibility conferred to that agent by the users of the social network and, consequently, the higher his/her influence on the other agents. Agents with high credibility are called \textit{influencers}: they are able to influence the opinion of other agents while being in turn virtually unaffected by the latter. Unlike~\cite{albi2017KRM}, in the present work we take the statistical distribution of the connectivity of the agents constant in time. This corresponds to assuming that the connectivity distribution possibly evolves over a time scale by far much slower than that of the opinion changes, so that it can be considered an almost stationary background.

Similarly to~\cite{toscani2006CMS}, individual changes of opinion result from random binary interactions between the agents, where now, in addition to the classical rules of change, the connectivity (viz. credibility) enters to modify the post-interaction opinions. In a microscopic binary interaction between two agents with states $(w,\,c)$ and $(w_\ast,\,c_\ast)$ the opinion variables update now according to 
\begin{equation}
    \begin{cases}
        w'=w+\gamma\kappa(c,\,c_\ast)(w_\ast-w)+D_\op(w,\,c)\eta \\
        w_\ast'=w_\ast+\gamma\kappa(c_\ast,\,c)(w-w_\ast)+D_\op(w_\ast,\,c_\ast)\eta_\ast.
    \end{cases}
    \label{eq:binary.w.1}
\end{equation}
In~\eqref{eq:binary.w.1}, $\gamma>0$ is a proportionality parameter. The function $\kappa$, expressing the rate of relaxation of either opinion toward that of the other agent (the compromise), is here depending on the connectivities of the agents. From the previous discussion about the role of connectivity, it is natural to require that $\kappa$ satisfies the assumptions:
$$ \kappa(c,\,c_\ast)\to 1 \quad \text{for} \quad \frac{c}{c_\ast}\to 0, \qquad
    \kappa(c,\,c_\ast)\to 0 \quad \text{for} \quad \frac{c}{c_\ast}\to +\infty, $$
which heuristically mean that the post-interaction opinion $w'$ is greatly influenced by $w_\ast$ if $c_\ast\gg c$ while it is virtually unaffected by $w_\ast$ if, conversely, $c_\ast\ll c$. Possible examples of functions with these characteristics are
\begin{align}
    \kappa(c,\,c_\ast) &= \frac{c_\ast}{c+c_\ast}=\frac{1}{1+{c}/{c_\ast}}, \nonumber \\
    \kappa(c,\,c_\ast) &= e^{-{c}/{c_\ast}}\left(1-e^{-{c_\ast}/{c}}\right) \label{eq:k}.
\end{align}
Note that both functions in~\eqref{eq:k} satisfy the further property $0\leq\kappa(c,\,c_\ast)\leq 1$ for all $c,\,c_\ast\in\R_+$.

The terms $D_\op(w,\,c)\eta$ and $D_\op(w_\ast,\,c_\ast)\eta_\ast$ measure the rate of change of opinion due to the self-thinking of the individuals, namely the possibility that agents change randomly (and independently of each other) their opinion. Following~\cite{toscani2006CMS}, we assume that $\eta$ and $\eta_\ast$ are independent and identically distributed random variables with zero mean and variance $\sigma^2>0$, while the coefficient $D_\op(w,\,c)=D_\op(\abs{w},\,c)$ is nonnegative for all $(w,\,c)\in [-1,\,1]\times\R_+$, non-increasing in $\abs{w}$ and vanishing for $\abs{w}\to 1$.
    
\subsection{Analysis of the microscopic interactions}
\label{sect:analysis.micro}
It is clear that, in order to have a physically acceptable model, the post-interaction opinions cannot cross the extremal values $\pm 1$. For this reason, it is important to verify that the binary interactions~\eqref{eq:binary.w.1} preserve the bounds, i.e., that $w',\,w_\ast'\in [-1,\,1]$ if $w,\,w_\ast\in [-1,\,1]$. Since $\abs{w_\ast}\leq 1$, we deduce from~\eqref{eq:binary.w.1} the bound
\begin{align*}
    \abs{w'} &= \abs{(1-\gamma\kappa(c,\,c_\ast))w+\gamma\kappa(c,\,c_\ast)w_\ast+D_\op(w,\,c)\eta} \\
    &\leq \abs{(1-\gamma\kappa(c,\,c_\ast)w}+\gamma\kappa(c,\,c_\ast)+D_\op(w,\,c)\abs{\eta}.
\intertext{If we further assume $0\leq \gamma,\,\kappa(c,\,c_\ast)\leq 1$,}
    \abs{w'}&\leq (1-\gamma\kappa(c,\,c_\ast))\abs{w}+\gamma\kappa(c,\,c_\ast)+D_\op(w,\,c)\abs{\eta}.
\end{align*}
Thus  $\abs{w'}\leq 1$ if $D_\op(w,\,c)\abs{\eta}\leq (1-\gamma\kappa(c,\,c_\ast))(1-\abs{w})$, which is satisfied if there exists a constant $\alpha>0$ such that
\begin{equation*}
    \begin{cases}
        \abs{\eta}\leq\alpha(1-\gamma\kappa(c,\,c_\ast)) \\[1mm]
        \alpha D_\op(w,\,c)\leq 1-\abs{w}.
    \end{cases}
\end{equation*}
The first condition can be enforced by requiring $\abs{\eta}\leq\alpha(1-\gamma)$, which gives a bound on the random variable $\eta$ independent of the connectivity distribution in the network. The second condition implies $D_\op(\pm 1,\,c)=0$ for all $c\in\R_+$, which characterizes the stubbornness of the agents with extremal opinions.

Next, denoting by $\ave{\cdot}$ the average with respect to the distributions of the random variables $\eta,\,\eta_\ast$, we observe that
$$ \ave{w'+w_\ast'}=w+w_\ast+\gamma(\kappa(c,\,c_\ast)-\kappa(c_\ast,\,c))(w_\ast-w). $$
Hence the mean opinion remains unchanged in a binary interaction only if the two agents have the same connectivity ($c=c_\ast$), or if the function $\kappa$ is symmetric, so that $\kappa(c,\,c_\ast)=\kappa(c_\ast,\,c)$.

Likewise, for $\gamma$ small we compute
\begin{align*}
    \ave{(w')^2+(w_\ast')^2} &= -2\gamma(\kappa(c_\ast,\,c)w_\ast-\kappa(c,\,c_\ast)w)(w_\ast-w) \\
    &\phantom{=} +\sigma^2(D_\op^2(w,\,c)+D_\op^2(w_\ast,\,c_\ast))+o(\gamma).
\end{align*}
We remark that, unlike the case of classical opinion dynamics (without social network)~\cite{toscani2006CMS}, the second moment (the energy) is not necessarily dissipated in a binary interaction, not even in the absence of self-thinking (i.e., for $\sigma^2=0$), unless again $\kappa$ is symmetric in its arguments.

\subsection{Boltzmann-type description}
\label{sect:Boltzmann}
Let us denote by $p=p(t,\,w,\,c)$ the proportion of agents in the population with opinion $w$ and connectivity $c$ at time $t\ge 0$. It is then possible to describe the time evolution of $p=p(t,\,w,\,c)$ by resorting to a Boltzmann-type equation, whose collision part reflects the dynamics of opinion changes because of the interactions among the agents. Under the binary rules~\eqref{eq:binary.w.1}, the Boltzmann-type equation reads
\begin{align}
    \begin{aligned}[b]
        \partial_tp &= Q_\op(p,\,p) \\
        &= \ave{\int_{\R_+}\int_{-1}^1\left(\frac{1}{\pr{J_\op}}p(t,\,\pr{w},\,c)p(t,\,\pr{w_\ast},\,c_\ast)
            -p(t,\,w,\,c)p(t,\,w_\ast,\,c_\ast)\right)\,dw_\ast\,dc_\ast},
    \end{aligned}
    \label{eq:Boltzmann.1.strong}
\end{align}
where $Q_\op$ is the collisional operator that takes into account opinion variations due to interactions, and the average $\langle \cdot \rangle$ is taken with respect to the distribution of the random variables $\eta$, $\eta_\ast$. In~\eqref{eq:Boltzmann.1.strong}, $\pr{w}$, $\pr{w_\ast}$ denote the pre-interaction opinions which generate the post-interaction opinions $w$, $w_\ast$ and $\pr{J_\op}$ is the Jacobian of the transformation~\eqref{eq:binary.w.1} as a function of the variables $\pr{w}$, $\pr{w_\ast}$. The weak form of this equation, which avoids the explicit computation of the Jacobian, writes
\begin{align}
    \begin{aligned}[b]
        \frac{d}{dt}&\int_{\R_+}\int_{-1}^1\phi(w,\,c)p(t,\,w,\,c)\,dw\,dc \\
        &=\ave{\int_{\R_+}\int_{-1}^1\int_{\R_+}\int_{-1}^1(\phi(w',\,c)-\phi(w,\,c))p(t,\,w,\,c)p(t,\,w_\ast,\,c_\ast)\,dw\,dc\,dw_\ast\,dc_\ast},
    \end{aligned}
    \label{eq:Boltzmann.1.weak}
\end{align}
where $\phi:[-1,\,1]\times\R_+\to\R$ is a test function, namely any observable quantity which can be expressed as a function of the microscopic state $(w,\,c)$, and $w'$ is the post-interaction opinion directly given by~\eqref{eq:binary.w.1}.

In order to simplify the kinetic description delivered by~\eqref{eq:Boltzmann.1.weak}, and to make it more amenable to mathematical analysis, we introduce the following argument. When picking randomly an agent of the system, we may assume that his/her opinion $w$ and connectivity $c$ are two independent variables. In other words, we assume that a generic microscopic state $(w,\,c)$ can be built by sampling independently the opinion $w$ from the statistical distribution of the opinions and the connectivity $c$ from the statistical distribution of the connectivity. In fact at the aggregate level there might be no \textit{a priori} reason to believe that a certain opinion is expressed only by individuals with a certain connectivity nor that a given connectivity level is possessed only by individuals expressing a certain opinion. Hence we postulate
\begin{equation}
    p(t,\,w,\,c)=f(t,\,w)g(c),
    \label{eq:p}
\end{equation}  
where $f(t,\,w)$ is the probability density function of the opinion at time $t$ and $g(c)$ is the probability density function of the connectivity. We anticipate that we will deal with a more general case later in Section~\ref{sect:p.gen}.

Plugging~\eqref{eq:p} into~\eqref{eq:Boltzmann.1.weak} and choosing the test function of the form $\phi(w,\,c)=\varphi(w)\psi(c)$ gives
\begin{align*}
    &\left(\int_{\R_+}\psi(c)g(c)\,dc\right)\frac{d}{dt}\int_{-1}^1\varphi(w)f(t,\,w)\,dw \\
    &\quad =\ave{\int_{\R_+}\int_{\R_+}\psi(c)\left(\int_{-1}^1\int_{-1}^1(\varphi(w')-\varphi(w))f(t,\,w)f(t,\,w_\ast)\,dw\,dw_\ast\right)
        g(c)g(c_\ast)\,dc\,dc_\ast}.
\end{align*}
In particular, for $\psi(c)\equiv 1$ we end up with
\begin{align}
    \begin{aligned}[b]
        \frac{d}{dt}&\int_{-1}^1\varphi(w)f(t,\,w)\,dw \\
        &= \ave{\int_{\R_+}\int_{\R_+}\left(\int_{-1}^1\int_{-1}^1(\varphi(w')-\varphi(w))f(t,\,w)f(t,\,w_\ast)\,dw\,dw_\ast\right)
            g(c)g(c_\ast)\,dc\,dc_\ast}.
    \end{aligned}
    \label{eq:Boltzmann.2}
\end{align}

Now we can take advantage of~\eqref{eq:Boltzmann.2} to investigate the aggregate trend of the mean opinion and of the energy of the agent's system.

Let $m(t):=\int_{-1}^1wf(t,\,w)\,dw$ be the mean opinion at time $t$. Choosing $\varphi(w)=w$ in~\eqref{eq:Boltzmann.2} we get
$$ \frac{dm}{dt}=\gamma\int_{\R_+}\int_{\R_+}\kappa(c,\,c_\ast)
    \left(\int_{-1}^1\int_{-1}^1(w_\ast-w)f(t,\,w)f(t,\,w_\ast)\,dw\,dw_\ast\right)g(c)g(c_\ast)\,dc\,dc_\ast=0, $$
therefore $m$ is conserved in time in spite of the fact that, as seen in Section~\ref{sect:analysis.micro}, at the level of a single binary interaction it is not.

In order to study the large time trend of the opinion energy, which will provide insights into the convergence of the system toward equilibria, it is convenient to apply to~\eqref{eq:Boltzmann.2} the \textit{quasi-invariant opinion limit} introduced in~\cite{toscani2006CMS}, namely an asymptotic procedure reminiscent of the grazing collision limit of the classical kinetic theory~\cite{villani1998ARMA}. In practice, we consider the regime of weak but frequent binary interactions, which amounts to taking $\gamma,\,\sigma^2\to 0^+$ in~\eqref{eq:binary.w.1} while simultaneously scaling the time as $\tau:=\gamma t$. Introducing the scaled distribution function $\tilde{f}(\tau,\,w):=f(\tau/\gamma,\,w)$, we easily obtain from~\eqref{eq:Boltzmann.2}
\begin{align}
    \begin{aligned}[b]
        \frac{d}{d\tau}&\int_{-1}^1\varphi(w)\tilde{f}(\tau,\,w)\,dw \\
        &= \frac{1}{\gamma}\ave{\int_{\R_+}\int_{\R_+}\left(\int_{-1}^1\int_{-1}^1(\varphi(w')-\varphi(w))
            \tilde{f}(\tau,\,w)\tilde{f}(\tau,\,w_\ast)\,dw\,dw_\ast\right)g(c)g(c_\ast)\,dc\,dc_\ast}.
    \end{aligned}
    \label{eq:Boltzmann.3}
\end{align}
Notice that for $\gamma$ small $t=\tau/\gamma$ is large. Therefore, for every fixed $\tau>0$ the limit $\gamma\to 0^+$ describes the large time trend of $f(t,\,w)$ and of its statistical moments. In parallel, since for $\tau\to+\infty$ it results $t\to+\infty$ as well, the asymptotic behavior of $\tilde{f}(\tau,\,w)$ and of its statistical moments approximates well that of $f(t,\,w)$ and of the corresponding statistical moments. If we define the opinion energy as $E(\tau):=\int_{-1}^1w^2\tilde{f}(\tau,\,w)\,dw$ and choose $\varphi(w)=w^2$ in~\eqref{eq:Boltzmann.3} we get, in the quasi-invariant opinion limit $\gamma,\,\sigma^2\to 0^+$,
\begin{equation}
    \frac{dE}{d\tau}=\cC(m^2-E)
        +\lim_{\gamma,\,\sigma^2\to 0^+}\frac{\sigma^2}{\gamma}\int_{\R_+}\int_{-1}^1D_\op^2(w,\,c)\tilde{f}(\tau,\,w)g(c)\,dw\,dc
    \label{eq:E}
\end{equation}
where
\begin{equation}
    \cC:=2\int_{\R_+}\int_{\R_+}\kappa(c,\,c_\ast)g(c)g(c_\ast)\,dc\,dc_\ast.
    \label{eq:C}
\end{equation}

If $\sigma^2/\gamma\to 0$ then from~\eqref{eq:E} we deduce
$$ E(\tau)=(E_0-m^2)e^{-\cC\tau}+m^2, $$
where $E_0:=E(0)$. Hence the energy converges exponentially fast to the asymptotic value $E^\infty=m^2$. The coefficient $\cC$, which depends on the statistical properties of the connectivity of the social network, gives the speed of convergence of $E$ to $E^\infty$. In other words, $1/\cC$ is proportional to the half-life of the exponential decay of the energy. Moreover, it is straightforward to see that
$$ W_2(\tilde{f}(\tau,\,\cdot),\,\delta_m)=\sqrt{E(\tau)-m^2}, $$
where $W_2$ at the left-hand side denotes the $2$-Wasserstein distance in the space of the probability measures, cf.~\cite{ambrosio2008BOOK}, and $\delta_m$ is the Dirac delta centered at the mean opinion $m$. Hence we further get
$$ W_2(\tilde{f}(\tau,\,\cdot),\,\delta_m)=\sqrt{E_0-m^2}\,e^{-\frac{\cC}{2}\tau}, $$
indicating that $\tilde{f}(\tau,\,w)$ converges asymptotically to $\delta_m$ (\textit{consensus}) with an exponential speed determined by the background network through $\cC/2$. Remarkably, in this case the energy is globally dissipated although it is not necessarily so in each binary interaction (cf. Section~\ref{sect:analysis.micro}).

If instead $\sigma^2/\gamma\to\lambda>0$ then from~\eqref{eq:E} we have
$$ \frac{dE}{d\tau}=\cC(m^2-E)+\lambda\int_{\R_+}\int_{-1}^1D_\op^2(w,\,c)\tilde{f}(\tau,\,w)g(c)\,dw\,dc. $$
Choosing for instance
\begin{equation}
    D_\op(w,\,c)=\beta(c)\sqrt{1-w^2}
    \label{eq:D}
\end{equation}
with $\beta(c)\geq 0$ for all $c\in\R_+$, and setting
\begin{equation}
    \cB:=\int_{\R_+}\beta^2(c)g(c)\,dc,
    \label{eq:B}
\end{equation}
the equation of the energy becomes
$$ \frac{dE}{d\tau}=\cC(m^2-E)+\lambda\cB(1-E), $$
whose solution reads
$$ E(\tau)=E_0e^{-(\cC+\lambda\cB)\tau}+\frac{\cC m^2+\lambda\cB}{\cC+\lambda\cB}\left(1-e^{-(\cC+\lambda\cB)\tau}\right). $$
The asymptotic value is now $E^\infty:=\frac{\cC m^2+\lambda\cB}{\cC+\lambda\cB}$ and the exponential speed at which it is approached depends on the background network through $\cC+\lambda\cB$ or, in other words, the half-life of the exponential decay is proportional to $1/(\cC+\lambda\cB)$. Since $\lambda,\,\cB\geq 0$, this speed is in general higher than the one of the case without self-thinking.

For a more accurate characterization of the asymptotic distribution reached in this case we refer the reader to the next Section~\ref{sect:FP}.

\begin{remark}
Let us define
$$ \norm{\kappa}:=\esssup_{(c,\,c_\ast)\in\R_+\times\R_+}\kappa(c,\,c_\ast). $$
Then, definition~\eqref{eq:C} implies that $\cC\leq 2\norm{\kappa}$. In particular, if $\kappa(c,\,c_\ast)\leq 1$ for all $c,\,c_\ast\in\R_+$ then $\cC\leq 2$. Let us observe that, in absence of the social network, the binary interaction rules  are often of the form~\eqref{eq:binary.w.1} with $\kappa\equiv 1$ (cf.~\cite{toscani2006CMS}), which obviously gives $\cC=2$. Here ``without social network'' means that in an opinion exchange there is no hierarchy among the agents based on their connectivity, viz. their credibility. In this case, convergence of the system toward the steady state is the fastest possible one. In other words, the presence of the network slows down this process.
\end{remark}

\subsection{Fokker-Planck asymptotic analysis with self-thinking}
\label{sect:FP}
In order to gain more detailed insights into the asymptotic opinion distribution reached by the system in the balanced interactive-diffusive regime $\sigma^2/\gamma\to\lambda>0$ we perform the quasi-invariant opinion limit in~\eqref{eq:Boltzmann.3} with a generic test function $\varphi$, which now we require to be smooth enough, say $\varphi\in C^3(-1,\,1)$, and such that $\varphi(\pm 1)=\varphi'(\pm 1)=0$.

Since $\gamma$, $\sigma^2$ are taken small, from~\eqref{eq:binary.w.1} we have that $w'-w$ is small as well and we can expand it in Taylor series to obtain
$$ \varphi(w')-\varphi(w)=\varphi'(w)(w'-w)+\frac{1}{2}\varphi''(w)(w'-w)^2+\frac{1}{6}\varphi'''(\bar{w})(w'-w)^3 $$
with $\bar{w}\in(\min\{w,\,w'\},\,\max\{w,\,w'\})$. Plugging this expansion into~\eqref{eq:Boltzmann.3} and developing the computations yields
\begin{align*}
    \frac{d}{dt}&\int_{-1}^1\varphi(w)\tilde{f}(\tau,\,w)\,dw \\
    &= \frac{\cC}{2}\int_{-1}^1\int_{-1}^1\varphi'(w)(w_\ast-w)\tilde{f}(\tau,\,w)\tilde{f}(\tau,\,w_\ast)\,dw\,dw_\ast \\
    &\phantom{=} +\frac{\gamma}{2}\left(\int_{\R_+}\int_{\R_+}\kappa^2(c,\,c_\ast)g(c)g(c_\ast)\,dc\,dc_\ast\right)
        \int_{-1}^1\int_{-1}^1\varphi''(w)(w_\ast-w)^2\tilde{f}(\tau,\,w)\tilde{f}(\tau,\,w_\ast)\,dw\,dw_\ast \\
    &\phantom{=} +\frac{\sigma^2}{2\gamma}\int_{\R_+}\int_{-1}^1\varphi''(w)D_\op^2(w,\,c)\tilde{f}(\tau,\,w)g(c)\,dw\,dc \\
    &\phantom{=} +R(\varphi),
\end{align*}
where the coefficient $\cC$ has been defined in~\eqref{eq:C} while the remainder is
$$ R(\varphi):=\frac{1}{6\gamma}\ave{\int_{\R_+}\int_{\R_+}\int_{-1}^1\int_{-1}^1\varphi'''(\bar{w})\left(\gamma\kappa(w_\ast-w)
    +D_\op\eta\right)^3\tilde{f}(w)\tilde{f}(w_\ast)g(c)g(c_\ast)\,dw\,dw_\ast\,dc\,dc_\ast}. $$
If $\langle\abs{\eta}^3\rangle<+\infty$ then $\langle\abs{\eta}^3\rangle\sim (\sigma^2)^{3/2}$; due to the further boundedness of $\varphi'''$, $\kappa$, $D_\op$, it results $\abs{R(\varphi)}\sim\gamma+\sqrt{\sigma^2}$, whence $R(\varphi)\to 0$ for $\gamma,\,\sigma^2\to 0^+$. On the whole, in the quasi-invariant opinion limit the previous equation becomes
\begin{align*}
    \frac{d}{dt}\int_{-1}^1\varphi(w)\tilde{f}(\tau,\,w)\,dw &=
        \frac{\cC}{2}\int_{-1}^1\int_{-1}^1\varphi'(w)(w_\ast-w)\tilde{f}(\tau,\,w)\tilde{f}(\tau,\,w_\ast)\,dw\,dw_\ast \\
    &\phantom{=} +\frac{\lambda}{2}\int_{\R_+}\int_{-1}^1\varphi''(w)D_\op^2(w,\,c)\tilde{f}(\tau,\,w)g(c)\,dw\,dc,
\end{align*}
which, integrating back by parts at the right-hand side and using the boundary conditions on $\varphi$, can be recognized as a weak form of the Fokker-Planck equation
\begin{equation}
    \partial_t\tilde{f}=\frac{\lambda}{2}\partial^2_w(\cD_\op(w)\tilde{f})+\frac{\cC}{2}\partial_w((w-m)\tilde{f}),
    \label{eq:FP.1}
\end{equation}
where we have defined
$$ \cD_\op(w):=\int_{\R_+}D_\op^2(w,\,c)g(c)\,dc. $$
The asymptotic solution $\tilde{f}^\infty(w):=\lim_{\tau\to+\infty}\tilde{f}(\tau,\,w)$ to this equation, which is obtained by equating the right-hand side of~\eqref{eq:FP.1} to zero, reads
$$ \tilde{f}^\infty(w)=\frac{K}{\cD_\op(w)}\exp{\left(\frac{\cC}{\lambda}\int\frac{m-w}{\cD_\op(w)}\,dw\right)}, $$
where $K>0$ is a normalization constant to be chosen in such a way that $\int_{-1}^1\tilde{f}^\infty(w)\,dw=1$. With $D_\op(w,\,c)$ given in particular by~\eqref{eq:D} we get $\cD_\op(w)=\cB(1-w^2)$, where the coefficient $\cB$ has been defined in~\eqref{eq:B}, and explicitly
\begin{equation}
    \tilde{f}^\infty(w)=\frac{(1+w)^{\frac{\cC(1+m)}{2\lambda\cB}-1}(1-w)^{\frac{\cC(1-m)}{2\lambda\cB}-1}}
        {2^{\frac{\cC}{\lambda\cB}-1}\Beta\left(\frac{\cC(1+m)}{2\lambda\cB},\,\frac{\cC(1-m)}{2\lambda\cB}\right)},
    \label{eq:finf.beta.1}
\end{equation}
where $\Beta(x,\,y):=\int_0^1t^{x-1}(1-t)^{y-1}\,dt$ is the beta function.

\begin{figure}[!t]
\centering
\includegraphics[scale=1]{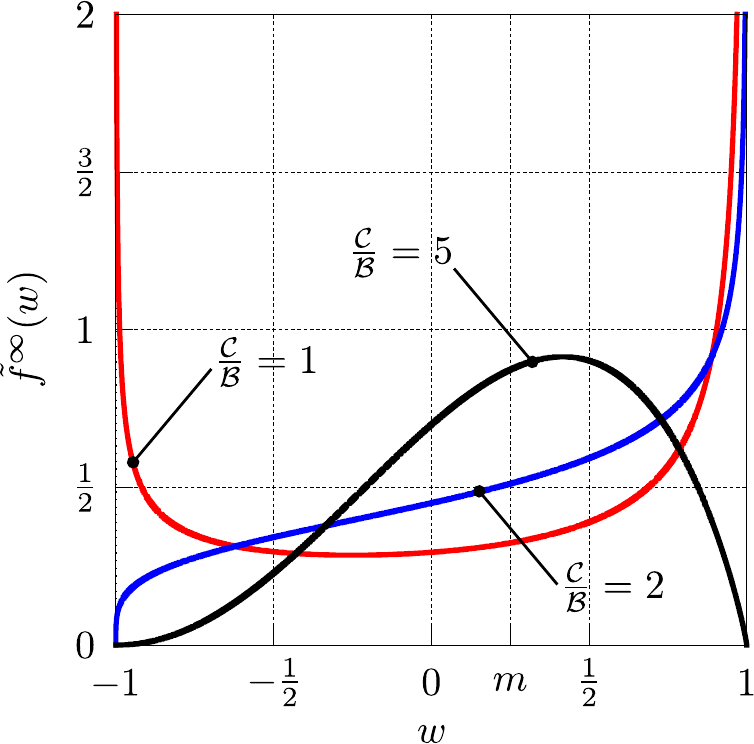}
\caption{The asymptotic distribution~\eqref{eq:finf.beta.1} with $m=\frac{1}{4}$, $\lambda=1$ and different values of $\frac{\cC}{\cB}$ representative of the three cases discussed in the text.}
\label{fig:beta}
\end{figure}

The function in~\eqref{eq:finf.beta.1} is a beta probability density function on the interval $[-1,\,1]$ parametrized by the constants $m\in [-1,\,1]$, $\lambda>0$, $\frac{\cC}{\cB}>0$. Its mean is $m$ and its variance is $\frac{1-m^2}{\frac{\cC}{\lambda\cB}+1}$. It is interesting to discuss the different trends of $\tilde{f}^\infty$ depending on the parameter $\frac{\cC}{\cB}$, which summarizes the statistical connectivity properties of the social network. By inspecting the exponents of the expression~\eqref{eq:finf.beta.1} we conclude in particular that:
\begin{enumerate}[label=(\roman*)]
\item if $\frac{\cC}{\cB}\geq\max\left\{\frac{2\lambda}{1+m},\,\frac{2\lambda}{1-m}\right\}$ then opinions mostly distribute around the mean $m$. This can be interpreted as a tendency to a mild consensus (cf. the black curve in Figure~\ref{fig:beta});
\item if $\min\left\{\frac{2\lambda}{1-m},\,\frac{2\lambda}{1+m}\right\}\leq\frac{\cC}{\cB}<\max\left\{\frac{2\lambda}{1-m},\,\frac{2\lambda}{1+m}\right\}$ then opinions tend to concentrate in $w=-1$ if $m<0$ or in $w=1$ if $m>0$, which indicates a tendency to a radicalized consensus (cf. the blue curve in Figure~\ref{fig:beta});
\item if $\frac{\cC}{\cB}<\min\left\{\frac{2\lambda}{1-m},\,\frac{2\lambda}{1+m}\right\}$ then opinions tend to radicalize at the two extreme values $w=\pm 1$ with a substantial splitting of the population between them (cf. the red curve in Figure~\ref{fig:beta}).
\end{enumerate}
We remark that by replacing $\sqrt{1-w^2}$ in~\eqref{eq:D} with a few different functions of $w$, such as $1-w^2$ or $1-\abs{w}$, different families of asymptotic distributions $\tilde{f}^\infty$ can be obtained analytically, cf.~\cite{toscani2006CMS}, which are still parametrized by the coefficient $\frac{\cC}{\cB}$. Then a similar analysis of the influence of the social network on the steady opinion distribution can be performed also in those cases. Here we stick to the choice~\eqref{eq:D} because it gives rise to an explicit closed form of relevant statistical moments of $\tilde{f}^\infty$.

\subsection{A more general case}
\label{sect:p.gen}
As mentioned in Section~\ref{sect:Boltzmann},  in this Section we will investigate the more general case in which we do not make the assumption of statistical independence of the variables $w$ and $c$, leading to~\eqref{eq:p}. To this purpose, and to avoid inessential difficulties, we slightly modify the interaction rules~\eqref{eq:binary.w.1} by assuming that the function $\kappa$ would depend only on the connectivity of the individual who is changing opinion, hence:
\begin{equation}
    \begin{cases}
        w'=w+\gamma\kappa(c)(w_\ast-w)+D_\op(w,\,c)\eta \\
        w_\ast'=w_\ast+\gamma\kappa(c_\ast)(w-w_\ast)+D_\op(w_\ast,\,c_\ast)\eta_\ast.
    \end{cases}
    \label{eq:binary.w.2}
\end{equation}
This simplification, which maintains most of the properties of the post-interaction opinions, allows us to explore this case with enough analytical detail. To this extent, we have to require that $\kappa$ be nonnegative and bounded. Moreover
$$ \lim_{c\to 0^+}\kappa(c)=1, \qquad \lim_{c\to+\infty}\kappa(c)=0. $$
This translates the idea that users of the social network with low connectivity, viz. with scarce credibility, may be more influenceable by possibly different opinions while those with high connectivity may tend to stick to their guns (sometimes they may need to do that in order to maintain the consensus).

The Boltzmann equation~\eqref{eq:Boltzmann.1.weak} with interaction rules~\eqref{eq:binary.w.2} and without assumption~\eqref{eq:p} is such that the mean opinion $m(t):=\int_{\R_+}\int_{-1}^1wp(t,\,w,\,c)\,dw\,dc$ is in general not conserved. In fact, letting $\phi(w,\,c)=w$ in~\eqref{eq:Boltzmann.1.weak} yields:
\begin{equation}
    \frac{dm}{dt}=\gamma\left(\int_{\R_+}\int_{-1}^1\kappa(c)p(t,\,w,\,c)\,dw\,dc\right)m
        -\gamma\int_{\R_+}\int_{-1}^1\kappa(c)wp(t,\,w,\,c)\,dw\,dc.
    \label{eq:m}
\end{equation}
In particular, denoting by $p^\infty(w,\,c):=\lim_{t\to+\infty}p(t,\,w,\,c)$ the steady distribution, it results that the stationary value $m^\infty$ of the mean opinion is formally given by
$$ m^\infty=\frac{\int_{\R_+}\int_{-1}^1w\, \kappa(c)\,p^\infty(w,\,c)\,dw\,dc}{\int_{\R_+}\int_{-1}^1\kappa(c)\,p^\infty(w,\,c)\,dw\,dc}. $$

We can make this formula more expressive by writing $p(t,\,w,\,c)=f_c(t,\,w)g(c)$, where $f_c(t,\,w)$ is the \textit{conditional distribution of $w$ given $c$}, whereas $g(c)$ is the same as in~\eqref{eq:p}. Notice that $f_c(t,\,w)$ is in general different from $f(t,\,w)$ in~\eqref{eq:p}, equality holding if and only if $w$ and $c$ are taken independent. Asymptotically  $p^\infty(w,\,c)=f^\infty_c(w)g(c)$, which implies
\begin{equation}
    m^\infty=\frac{\int_{\R_+}\kappa(c)\left(\int_{-1}^1wf_c^\infty(w)\,dw\right)g(c)\,dc}{\int_{\R_+}\kappa(c)g(c)\,dc}
        =\frac{\int_{\R_+}\kappa(c)m_c^\infty g(c)\,dc}{\int_{\R_+}\kappa(c)g(c)\,dc},
    \label{eq:minf.1}
\end{equation}
$m^\infty_c:=\int_{-1}^1wf_c^\infty(w)\,dw$ being the mean of the conditional distribution $f_c^\infty$.

By applying the quasi-invariant opinion limit to~\eqref{eq:Boltzmann.1.weak} with interaction rules~\eqref{eq:binary.w.2}, and proceeding like in Section~\ref{sect:FP}, one obtains the Fokker-Planck equation for the scaled distribution function $\tilde{p}(\tau,\,w,\,c):=p(\tau/\gamma,\,w,\,c)$ in the large time scale $\tau=\gamma t$, which is now given by
\begin{equation}
    \partial_\tau\tilde{p}=\frac{\lambda}{2}\partial_w^2(D_\op^2(w,\,c)\tilde{p})+\partial_w\bigl(\kappa(c)(w-m(\tau))\tilde{p}\bigr).
    \label{eq:FP.p}
\end{equation}
In equation~\eqref{eq:FP.p} the constant $\lambda>0$ is, as before, the limit of the ratio $\sigma^2/\gamma$ when $\gamma,\,\sigma^2\to 0^+$. Notice that, in spite of the fact that $\tilde{p}$ depends on both $w$ and $c$, this equation involves only $w$-derivatives (besides the time $\tau$) because the distribution of the connectivity is constant by assumption. Setting $\tilde{p}(\tau,\,w,\,c)=\tilde{f}_c(\tau,\,w)g(c)$, with $\tilde{f}_c(\tau,\,w)=f_c(\tau/\gamma,\,w)$,~\eqref{eq:FP.p} turns out to be actually an equation for $\tilde{f}_c$:
$$ \partial_\tau\tilde{f}_c=\frac{\lambda}{2}\partial_w^2(D_\op^2(w,\,c)\tilde{f}_c)+\partial_w\bigl(\kappa(c)(w-m(\tau))\tilde{f}_c\bigr), $$
whose stationary solution $\tilde{f}_c^\infty(w):=\lim_{\tau\to+\infty}\tilde{f}_c(\tau,\,w)$ satisfies the first-order differential equation
$$ \frac{\lambda}{2}\partial_w^2\bigl(D_\op^2(w,\,c)\tilde{f}_c^\infty\bigr)+\partial_w\bigl(\kappa(c)(w-m^\infty)\tilde{f}_c^\infty\bigr)=0. $$
Although $m^\infty$ is in general unknown, we can solve this equation by regarding it as a constant which parametrizes the solution $\tilde{f}_c^\infty$. We have:
$$ \tilde{f}^\infty_c(w)=\frac{K}{D_\op^2(w,\,c)}\exp{\left(\frac{2\kappa(c)}{\lambda}\int\frac{m^\infty-w}{D_\op^2(w,\,c)}\,dw\right)}, $$
where the two constants $K>0$, $m^\infty\in [-1,\,1]$ are determined by imposing
\begin{equation}
    \begin{cases}
        \displaystyle{\int_{-1}^1}\tilde{f}_c^\infty(w)\,dw=1
            & \text{($\tilde{f}_c^\infty$ is a probability density w.r.t. $w$ for all $c\in\R_+$)} \\
        \displaystyle{\int_{\R_+}\int_{-1}^1}\tilde{f}_c^\infty(w)g(c)\,dw\,dc=m^\infty
            & \text{($m^\infty$ is the mean of $\tilde{p}^\infty(w,\,c)=\tilde{f}_c^\infty(w)g(c)$)}.
    \end{cases}
    \label{eq:minf.2}
\end{equation}
The second condition may be equivalently replaced by~\eqref{eq:minf.1}.

Choosing the diffusion coefficient as in~\eqref{eq:D} we obtain 
\begin{equation}
    \tilde{f}_c^\infty(w)=\frac{(1+w)^{\frac{\kappa(c)\left(1+m^\infty\right)}{\lambda\beta^2(c)}-1}
        (1-w)^{\frac{\kappa(c)\left(1-m^\infty\right)}{\lambda\beta^2(c)}-1}}{2^{\frac{2\kappa(c)}{\lambda\beta^2(c)}-1}
            \Beta\left(\frac{\kappa(c)(1+m^\infty)}{\lambda\beta^2(c)},\,\frac{\kappa(c)(1-m^\infty)}{\lambda\beta^2(c)}\right)},
    \label{eq:finf.beta.2}
\end{equation}
which shows that, for every $c\in\R_+$, the steady distribution is a beta probability density function on the interval $[-1,\,1]$. Here again $\Beta$ denotes the beta function. Interestingly, the mean of~\eqref{eq:finf.beta.2} coincides precisely with $m^\infty$, thus it is in particular independent of $c$. Hence in this case conditions~\eqref{eq:minf.2} are automatically satisfied for every choice of $m^\infty\in [-1,\,1]$. This result can be fruitfully commented.
\begin{enumerate}[label=(\roman*)]
\item In general, studying the time-asymptotic problem may not be enough to identify univocally the stationary distributions. For instance, when the steady state is given as in~\eqref{eq:finf.beta.2} the value of $m^\infty$ needs to be determined from the time-evolutionary problem~\eqref{eq:m}, and this requires the knowledge of $p(t,\,w,\,c)$ from~\eqref{eq:Boltzmann.1.weak} for all $t>0$. However, for different choices of the diffusion coefficient $D_\op(w,\,c)$, such as e.g., those inspired by~\cite{toscani2006CMS}, conditions~\eqref{eq:minf.2} may give rise to a non-linear equation for $m^\infty$ which may admit solutions (though not necessarily unique). This method is similar to the one proposed in~\cite{visconti2017MMS} for recovering the stationary fundamental diagrams of vehicular traffic from a kinetic approach.
\item If the initial distribution $p_0(w,\,c):=p(0,\,w,\,c)=f_{c,0}(w)g(c)$, with $f_{c,0}(w):=f_c(0,\,w)$, is chosen in such a way that the mean $m_c(0):=\int_{-1}^1wf_{c,0}(w)\,dw$ is independent of $c$, then, under the ansatz that also $m_c(t):=\int_{-1}^1wf_c(t,\,w)\,dw$ is independent of $c$ for every $t>0$, it follows that $m_c(t)$ coincides actually with the $w$-mean $m(t)$ of $p$ and further, from~\eqref{eq:m}, that this value is conserved in time. Hence, similarly to~\eqref{eq:finf.beta.1}, $m^\infty=m(0)$ in~\eqref{eq:finf.beta.2},  with the significant difference that, unlike $\tilde{f}^\infty$, here $\tilde{f}_c^\infty$ is not the $w$-marginal distribution of $p^\infty$ or, in other words, that it is not the asymptotic opinion distribution. The latter is instead given by
$$ \tilde{f}^\infty(w)=\int_{\R_+}\tilde{f}^\infty_c(w)g(c)\,dc. $$
\end{enumerate}

\section{Spreading of the popularity of a product}
\label{sect:popularity}
In this section we consider the problem of the \textit{spreading of the popularity} of a certain product induced by the opinion dynamics described by the model of Section~\ref{sect:opinion.dynamics}. The ``product'' may be any piece of information reaching the users of the social network: news, videos, advertisements, and the like, that individuals possibly repost to their followers depending on how much their current opinion is aligned with it.

\subsection{Microscopic model}
We quantify the popularity of a product by means of a variable $v\in\R_+$, whose evolution depends on the interaction with the opinion $w$ and the connectivity $c$ of the agents that the product reaches. In particular, we propose the following microscopic update rule:
\begin{equation}
    v'=(1-\mu)v+P(w,\,c)+D_\pop(v)\xi,
    \label{eq:binary.v}
\end{equation}
where $\mu\in (0,\,1)$ is the natural decay rate of the popularity of a product which is not reposted, $P:[-1,\,1]\times\R_+\to\R_+$ is a function expressing the increase in popularity due to reposting, and finally $\xi$ is a random variable with zero mean and finite variance $\varsigma^2>0$ modeling a stochastic fluctuation of the popularity with popularity-dependent strength $D_\pop(v)\geq 0$. A possible choice for $P$ is
\begin{equation}
    P(w,\,c)=\nu c\mathbb{1}_{[0,\,\Delta]}(\abs{w-\hat{w}}),
    \label{eq:P}
\end{equation}
where $\nu,\,\Delta>0$ are parameters, $\mathbb{1}_{[0,\,\Delta]}$ is the characteristic function of the interval $[0,\,\Delta]$, and $\hat{w}\in [-1,\,1]$ is the \textit{target opinion}, i.e., the opinion that the product is mainly addressed at. Hence, according to model~\eqref{eq:P}, the increase in the popularity depends on whether an individual decides to repost the product, which happens if his/her opinion is closer than the threshold $\Delta$ to the target opinion. In such a case, the increase in popularity is proportional to the connectivity of the individual, viz. to the number of followers reached when reposting the product.

In order to guarantee from~\eqref{eq:binary.v} that $v'\geq 0$ we observe that, since $P(w,\,c)\geq 0$, it is enough to require $D_\pop(v)\xi\geq (\mu-1)v$, which is satisfied if $\xi\geq\mu-1$ and $D_\pop(v)\leq v$. Hence the stochastic fluctuation can take negative values provided the diffusion coefficient is not larger than $v$. The latter characteristic induces in particular $D_\pop(0)=0$. The most straightforward choice for $D_\pop$ is $D_\pop(v)=v$, which implies a larger and larger diffusion for increasing popularity. Other options are:
$$ D_\pop(v)=\min\{v,\,V_0\}, \qquad D_\pop(v)=\min\left\{v,\,\frac{V_0^2}{v}\right\}, $$
where $V_0>0$. The first function expresses a saturation of the diffusion coefficient for high popularity ($v>V_0$). The second function expresses instead a decay of the popularity fluctuations for highly popular products (again $v>V_0$).

\subsection{Boltzmann-type description}
\label{sect:Boltzmann.v}
We now implement a Boltzmann-type kinetic description of the microscopic dynamics~\eqref{eq:binary.v}, coupled with either~\eqref{eq:binary.w.1} or~\eqref{eq:binary.w.2}, by introducing the distribution function $h=h(t,\,v):\R_+\times\R_+\to\R_+$ such that $h(t,\,v)\,dv$ is the fraction of products with popularity in $[v,\,v+dv]$ at time $t$. If $p(t,\,w,\,c)$ is the distribution of the pair $(w,\,c)$ at time $t$ as introduced in Section~\ref{sect:Boltzmann} we have:
\begin{equation}
    \begin{cases}
        \partial_tp=Q_\op(p,\,p) \\
        \partial_th=Q_\pop(h,\,p),
    \end{cases}
    \label{eq:system_Boltzmann}
\end{equation}
where the collisional operator $Q_\op$ has been defined in~\eqref{eq:Boltzmann.1.strong} whereas $Q_\pop$ writes as follows:
$$ Q_\pop(h,\,p)(t,\,v)=\ave{\int_{\R_+}\int_{-1}^1\left(\dfrac{1}{\pr{J_\pop}}h(t,\,\pr{v})-h(t,\,v)\right)p(t,\,w,\,c)\,dw\,dc}. $$
Here $\pr{v}$ is the pre-interaction value of the popularity which generates the post-interaction value $v$ according to the transformation~\eqref{eq:binary.v} while $\pr{J_\pop}$ is the Jacobian of such a transformation as a function of the variable $\pr{v}$. In weak form, the second equation in~\eqref{eq:system_Boltzmann} reads
\begin{equation}
    \frac{d}{dt}\int_{\R_+}\varphi(v)h(t,\,v)\,dv=
        \ave{\int_{\R_+}\int_{-1}^1\int_{\R_+}\left(\varphi(v')-\varphi(v)\right)h(t,\,v)p(t,\,w,\,c)\,dv\,dw\,dc},
    \label{eq:Boltzmann.pop}
\end{equation}
where $\ave{\cdot}$ denotes here the average with respect to the distribution of the random variable $\xi$ in~\eqref{eq:binary.v} and $\varphi:\R_+\to\R$ is a test function, i.e., any observable quantity of the popularity $v$.

If we assume in particular that opinion dynamics are a much quicker process than the spreading of the popularity then in~\eqref{eq:Boltzmann.pop} we can replace $p$ with the asymptotic distribution $p^\infty(w,\,c)$ to get
\begin{equation}
    \frac{d}{dt}\int_{\R_+}\varphi(v)h(t,\,v)\,dv=
        \ave{\int_{\R_+}\int_{-1}^1\int_{\R_+}\left(\varphi(v')-\varphi(v)\right)h(t,\,v)p^\infty(w,\,c)\,dv\,dw\,dc}.
    \label{eq:Boltzmann.v}
\end{equation}

Choosing $\varphi(v)=1$ shows that $\int_{\R_+}h(t,\,v)\,dv$ is constant in time, hence if we choose an initial distribution $h_0(v):=h(0,\,v)$ which satisfies the normalization condition $\int_{\R_+}h_0(v)\,dv=1$ then $h(t,\,v)$ will be a probability density function for all $t>0$.

Moreover, denoting by $M(t):=\int_{\R_+}vh(t,\,v)\,dv$ the average of the popularity and letting $\varphi(v)=v$ in~\eqref{eq:Boltzmann.v} yields
$$ \frac{dM}{dt}=-\mu M+\cP, \qquad
    \cP:=\int_{\R_+}\int_{-1}^1P(w,\,c)p^\infty(w,\,c)\,dw\,dc, $$
that is
$$ M(t)=\left(M_0-\frac{\cP}{\mu}\right)e^{-\mu t}+\frac{\cP}{\mu} $$
with $M_0:=M(0)$. As a result, the mean popularity reaches asymptotically the value $M^\infty=\frac{\cP}{\mu}$, which depends on the statistical properties of the opinion distribution and of the social network through $\cP$.

Likewise, denoting by $\cE(t):=\int_{\R_+}v^2h(t,\,v)\,dv$ the energy of the distribution $h$, we can study the asymptotic behavior of $\cE$ by taking advantage of the quasi-invariant limit procedure illustrated in Sections~\ref{sect:Boltzmann},~\ref{sect:FP}. Precisely, let $0<\epsilon\ll 1$ be a small parameter and scale $\mu=\mu_0\epsilon$ in~\eqref{eq:binary.v} and $\nu=\nu_0\epsilon$ in~\eqref{eq:P} with $\mu_0,\,\nu_0>0$. Moreover, let $P(w,\,c)=\epsilon P_0(w,\,c)$ with $P_0(w,\,c)=\nu_0c\mathbb{1}_{[0,\,\Delta]}(\abs{w-\hat{w}})$. In the large time scale $\tau:=\epsilon t$ the scaled distribution function $\tilde{h}(\tau,\,v)=h(\tau/\epsilon,\,v)$ satisfies the equation
\begin{equation}
    \frac{d}{d\tau}\int_{\R_+}\varphi(v)\tilde{h}(\tau,\,v)\,dv=
        \frac{1}{\epsilon}\ave{\int_{\R_+}\int_{-1}^1\int_{\R_+}\left(\varphi(v')-\varphi(v)\right)\tilde{h}(\tau,\,v)p^\infty(w,\,c)\,dv\,dw\,dc}.
    \label{eq:Boltzmann.v.eps}
\end{equation}
Choosing $\varphi(v)=v^2$ and taking the quasi-invariant interaction limit $\epsilon,\,\varsigma^2\to 0^+$ (recall that $\varsigma^2=\ave{\xi^2}$) we obtain
$$ \frac{d\cE}{d\tau}=-2\mu_0\cE+2\cP_0M+\lim_{\epsilon,\,\varsigma^2\to 0^+}\frac{\varsigma^2}{\epsilon}\int_{\R_+}D_\pop^2(v)\tilde{h}(\tau,\,v)\,dv, $$
where we have denoted $\cP_0:=\int_{\R_+}\int_{-1}^1P_0(w,\,c)p^\infty(w,\,c)\,dw\,dc$.

If $\varsigma^2/\epsilon\to 0^+$ then this equation reduces to
$$ \frac{d\cE}{d\tau}=-2\mu_0\cE+2\cP_0M, $$
which implies for the energy the asymptotic value $\cE^\infty=\cP_0M^\infty/\mu_0$. Noticing that with the $\epsilon$-scaling introduced above it results $M^\infty=\cP_0/\mu_0$, we finally obtain $\cE^\infty=(\cP_0/\mu_0)^2=(M^\infty)^2$. This indicates that the asymptotic popularity distribution $\tilde{h}^\infty(v):=\lim_{\tau\to+\infty}\tilde{h}(\tau,\,v)$ has zero variance, hence $\tilde{h}^\infty(v)=\delta_{\cP_0/\mu_0}(v)$.

If conversely $\varsigma^2/\epsilon\to\zeta>0$ and if we choose in particular $D_\pop(v)=v$ then the previous equation gives
\begin{equation}
    \frac{d\cE}{d\tau}=(\zeta-2\mu_0)\cE+2\cP_0M,
    \label{eq:cE}
\end{equation}
which implies $\cE^\infty=\frac{2\cP_0^2}{\mu_0(2\mu_0-\zeta)}$. From the constraint $\cE^\infty\geq 0$ we deduce that this value is admissible only if $\zeta<2\mu_0$. Instead if $\zeta\geq 2\mu_0$, i.e., for a too strong diffusion in the limit,~\eqref{eq:cE} indicates that $\cE^\infty\to+\infty$, hence that $\tilde{h}^\infty$ has infinite variance. In any case, for $\zeta>0$ the asymptotic popularity distribution is no longer a Dirac delta, i.e., it does not collapse onto the asymptotic mean value $M^\infty$. We defer to the next Section~\ref{sect:FP.v} a more detailed analysis of the large time behavior in this case.

\subsection{Fokker-Planck asymptotic analysis with popularity diffusion}
\label{sect:FP.v}
We can apply to~\eqref{eq:Boltzmann.v.eps} the same quasi-invariant limit procedure discussed in Section~\ref{sect:FP}, taking $\epsilon,\,\varsigma^2\to 0^+$ and assuming $\varsigma^2/\epsilon\to\zeta>0$. This describes an asymptotic regime in which the effects of natural decay plus reposting and diffusion balance. The resulting Fokker-Planck equation for the distribution $\tilde{h}(\tau,\,v)$ is
$$ \partial_\tau\tilde{h}=\frac{\zeta}{2}\partial^2_v(D_\pop^2(v)\tilde{h})+\partial_v((\mu_0v-\cP_0)\tilde{h}), $$
which admits the following stationary solution:
$$ \tilde{h}^\infty(v)=\frac{K}{D_\pop^2(v)}\exp{\left(\frac{2}{\zeta}\int\frac{\cP_0-\mu_0v}{D_\pop^2(v)}\,dv\right)} $$
where $K>0$ is a normalization constant such that $\int_{\R_+}\tilde{h}^\infty(v)\,dv=1$.

\begin{figure}[!t]
\centering
\includegraphics[scale=1]{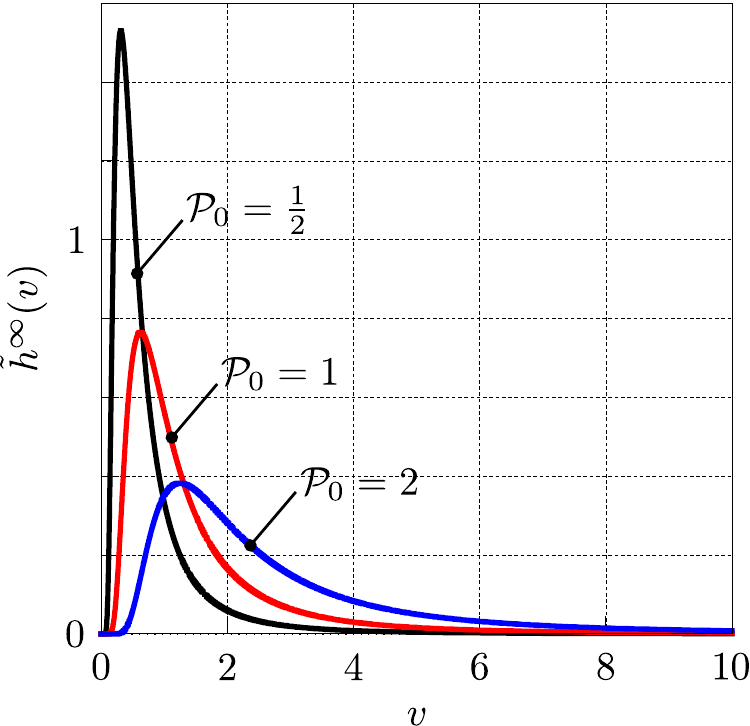}
\caption{The asymptotic distribution~\eqref{eq:h.Pareto} with $\mu_0=\frac{3}{5}$, $\zeta=1$ and different values of $\cP_0$.}
\label{fig:Pareto}
\end{figure}

For $D_\pop(v)=v$ we obtain in particular
\begin{equation}
    \tilde{h}^\infty(v)=\frac{(2\cP_0/\zeta)^{1+\frac{2\mu_0}{\zeta}}}{\Gamma\bigl(1+\frac{2\mu_0}{\zeta}\bigr)}\cdot
        \frac{e^{-\frac{2\cP_0}{\zeta v}}}{v^{2\left(1+\frac{\mu_0}{\zeta}\right)}},
    \label{eq:h.Pareto}
\end{equation}
where $\Gamma(z):=\int_0^{+\infty}t^{z-1}e^{-t}\,dt$ is the gamma function. Notice that~\eqref{eq:h.Pareto} is a fat-tailed inverse gamma distribution, indeed $\tilde{h}^\infty(v)\sim v^{-2\left(1+\frac{\mu_0}{\zeta}\right)}$ when $v\to +\infty$. Precisely, this distribution exhibits a \textit{Pareto tail}, cf.~\cite{gualandi2017ECONOMICS}, which indicates that products reaching very high popularity levels may be rare in general but not that improbable. The mean of the distribution~\eqref{eq:h.Pareto} is $\cP_0/\mu_0$, consistently with the result found in Section~\ref{sect:Boltzmann.v}. Notice also that $v^2\tilde{h}^\infty(v)\sim v^{-\frac{2\mu_0}{\zeta}}$ for $v\to +\infty$, which confirms that the energy and the variance of the distribution~\eqref{eq:h.Pareto} are finite only if $\zeta<2\mu_0$.

Figure~\ref{fig:Pareto} shows different profiles of the distribution~\eqref{eq:h.Pareto} for fixed $\mu_0,\,\zeta$ and increasing values of the parameter $\cP_0$, which accounts for the amount of popularity pumped into the system by the users of the social network through reposting.

For $D_\pop(v)=\min\{v,\,V_0\}$, $V_0>0$, the stationary distribution reads instead
\begin{equation}
    \tilde{h}^\infty(v)=
        \begin{cases}
            K\dfrac{e^{-\frac{2\cP_0}{\zeta v}}}{v^{2\left(1+\frac{\mu_0}{\zeta}\right)}} & \text{if } v\leq V_0 \\[5mm]
            \dfrac{Ke^{\frac{\mu_0}{\zeta}\left(1-\frac{4\cP_0}{\mu_0V_0}\right)}}{V_0^{2\left(1+\frac{\mu_0}{\zeta}\right)}}
                e^{\frac{2}{\zeta V_0^2}\left(\cP_0-\frac{\mu_0}{2}v\right)v} & \text{if } v>V_0.
        \end{cases}
    \label{eq:h.notfat.1}
\end{equation}
Notice that now $\tilde{h}^\infty(v)\sim e^{-\frac{\mu_0}{\zeta V_0^2}v^2}$ when $v\to +\infty$, thus fat tails disappear immediately if the diffusion coefficient is assumed to saturate at high popularity.

Finally, for $D_\pop(v)=\min\left\{v,\,\frac{V_0^2}{v}\right\}$, $V_0>0$, the stationary distribution becomes
\begin{equation}
    \tilde{h}^\infty(v)=
        \begin{cases}
            K\dfrac{e^{-\frac{2\cP_0}{\zeta v}}}{v^{2\left(1+\frac{\mu_0}{\zeta}\right)}} & \text{if } v\leq V_0 \\[5mm]
            \dfrac{Ke^{\frac{\mu_0}{2\zeta}\left(1-\frac{16\cP_0}{\mu_0V_0}\right)}}{V_0^{2\left(2+\frac{\mu_0}{\zeta}\right)}}
                v^2e^{\frac{2}{3\zeta V_0^4}\left(\cP_0-\frac{3\mu_0}{4}v\right)v^3} & \text{if } v>V_0,
        \end{cases}
    \label{eq:h.notfat.2}
\end{equation}
which has an even thinner tail than~\eqref{eq:h.notfat.1}, indeed $\tilde{h}^\infty(v)\sim v^2e^{-\frac{\mu_0}{2\zeta V_0^4}v^4}$ for $v\to +\infty$.

The distribution of popularity $\tilde{h}^\infty$ supported the most by empirical evidence is~\eqref{eq:h.Pareto}. For instance, in~\cite{redner1998EPJB} the author finds that the statistical distribution of the popularity of scientific papers, measured in terms of the received citations, features a power-law-type tail with Pareto exponent near to $3$. In~\cite{sinha2004EPJB} the authors find that the popularity of the movies in the United States, measured in terms of their box office gross income, exhibits completely analogous statistical properties. 

The Pareto exponent of the distribution~\eqref{eq:h.Pareto} can be computed out of the probability that the popularity be greater than a given threshold $v>0$:
$$ \frac{Ce^{-\frac{2\cP_0}{\zeta v}}}{v^{1+\frac{2\mu_0}{\zeta}}}\leq\int_v^{+\infty}\tilde{h}^\infty(u)\,du
    \leq\frac{C}{v^{1+\frac{2\mu_0}{\zeta}}}, $$
where $C>0$ is a constant. Hence $\int_v^{+\infty}\tilde{h}^\infty(u)\,du\sim v^{-\left(1+\frac{2\mu_0}{\zeta}\right)}$ when $v\to +\infty$, giving the Pareto exponent $1+\frac{2\mu_0}{\zeta}$. Owing to the constraint $\zeta<2\mu_0$, this exponent is invariably greater than $2$ and is near $3$ if $\mu_0/\zeta$ is near $1$, i.e., if the natural decay rate of popularity is of the same order as the variance of the stochastic fluctuations.

\section{Numerical experiments}
\label{sect:numerics}
This section is devoted to a numerical investigation of the kinetic models~\eqref{eq:Boltzmann.1.weak},~\eqref{eq:Boltzmann.pop} of the opinion dynamics over a background social network and of the resulting spreading of the popularity of products. The numerical approximation of the Boltzmann equations is done by means of Monte Carlo methods in the Fokker-Planck scaling, see~\cite{pareschi2001ESAIMP,pareschi2013BOOK,Pareschi2018}. In the following we use in particular samples of $N=10^5$ particles.

\subsection{Test 1: Opinion dynamics with independent \texorpdfstring{$\boldsymbol{w}$}{}, \texorpdfstring{$\boldsymbol{c}$}{}}
\label{sect:test1}
In this first test we consider the opinion dynamics model~\eqref{eq:binary.w.1}-\eqref{eq:Boltzmann.1.strong} under the independence assumption~\eqref{eq:p} between $w$ and $c$, which leads to equation~\eqref{eq:Boltzmann.2} for the evolution of the marginal opinion distribution $f(t,\,w)$. We remark that in this case the asymptotic profile of $f$ can be analytically computed in the quasi-invariant opinion regime, cf. Section~\ref{sect:FP} and in particular~\eqref{eq:finf.beta.1}. Therefore this test serves as a benchmark for successive numerical experiments.

We consider the following connectivity distributions for $c\in\R_+$:
\begin{equation}
    g_1(c)=\frac{e^{-\frac{1}{c}}}{c^3}, \qquad g_2(c)=e^{-c},
    \label{eq:g12}
\end{equation}
which are, respectively, a power law of degree $3$ (for $c$ large) and an exponential distribution. Notice that both distributions imply a unitary mean connectivity, indeed
$$ \int_{\R_+}cg_1(c)\,dc=\int_{\R_+}cg_2(c)\,dc=1. $$
The choice of $g_1$ in~\eqref{eq:g12} is motivated by the experimental literature, see e.g.,~\cite{barabasi1999SCIENCE,clauset2009SIREV,newman2002PNAS} among others, according to which the connectivity distribution of many networks resulting from social interactions, such as e.g., contact/follower networks in online platforms, scientific collaboration networks, economic networks, is of power-law type. This feature is commonly assumed to be generated by the tendency of the individuals to create links with highly connected vertices of the network rather than with poorly connected ones. Conversely the choice of $g_2$ in~\eqref{eq:g12} is motivated by the fact that many social networks feature the so-called \textit{small-world} structure, i.e., they are highly clustered with small characteristic path lengths~\cite{barabasi1999PHYSA,watts1998NATURE}.

For both networks modeled by $g_1$ and $g_2$ we compute numerically the large time trend of the solution $f$ to the Boltzmann equation~\eqref{eq:Boltzmann.2} using in the binary interaction rules~\eqref{eq:binary.w.1} the functions $\kappa$ given by~\eqref{eq:k} and $D_\op$ given by~\eqref{eq:D} with in particular
\begin{equation}
    \beta(c)=\frac{1}{\frac{1}{10}+c}.
    \label{eq:test1.beta}
\end{equation}
This function models a local diffusion which becomes weaker and weaker as the connectivity increases, meaning that people with higher credibility tend to be less prone to erratic changes of opinions due to self-thinking.

\begin{figure}[!t]
\includegraphics[scale=0.5]{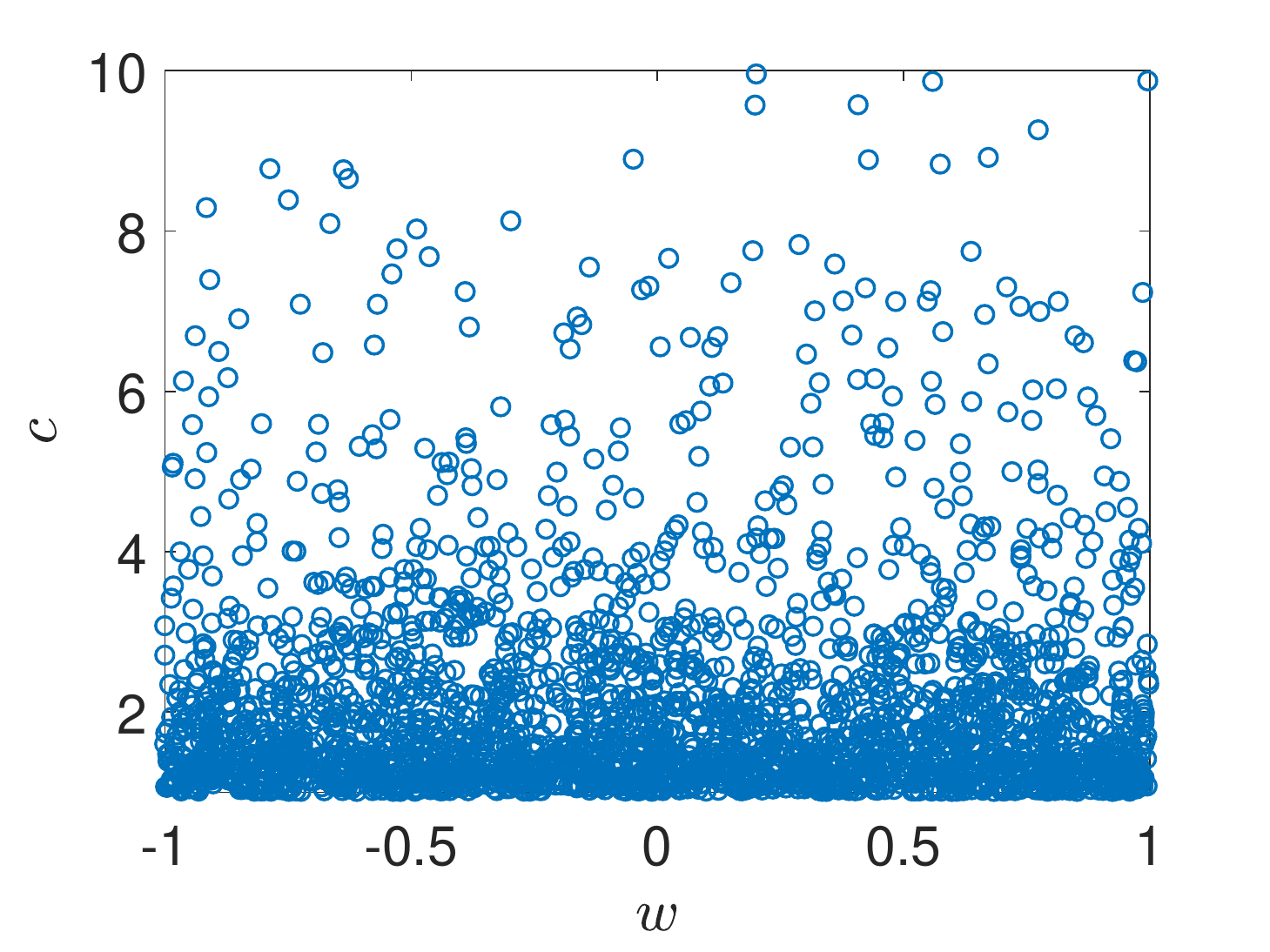}
\includegraphics[scale=0.5]{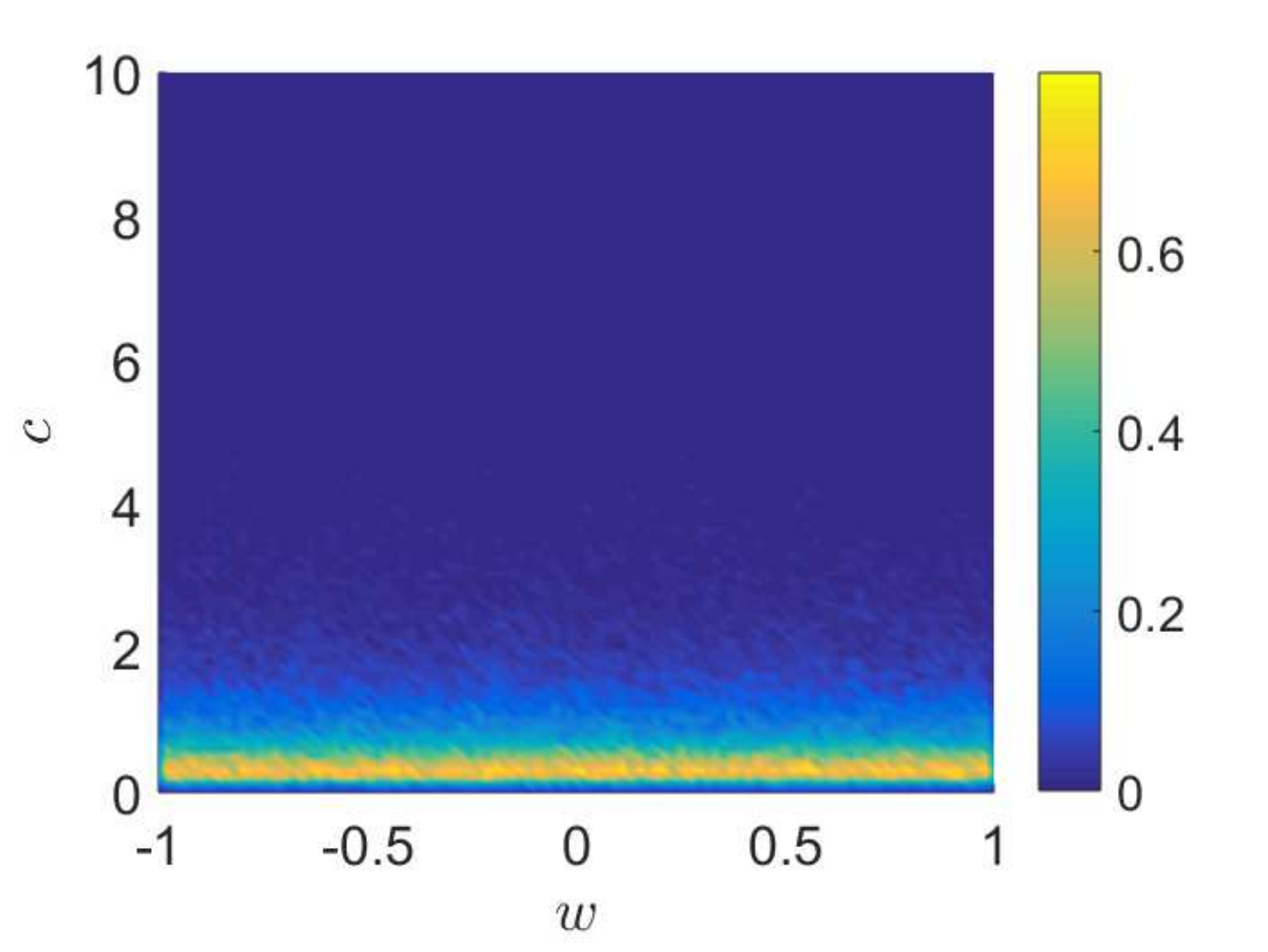} \\
\includegraphics[scale=0.5]{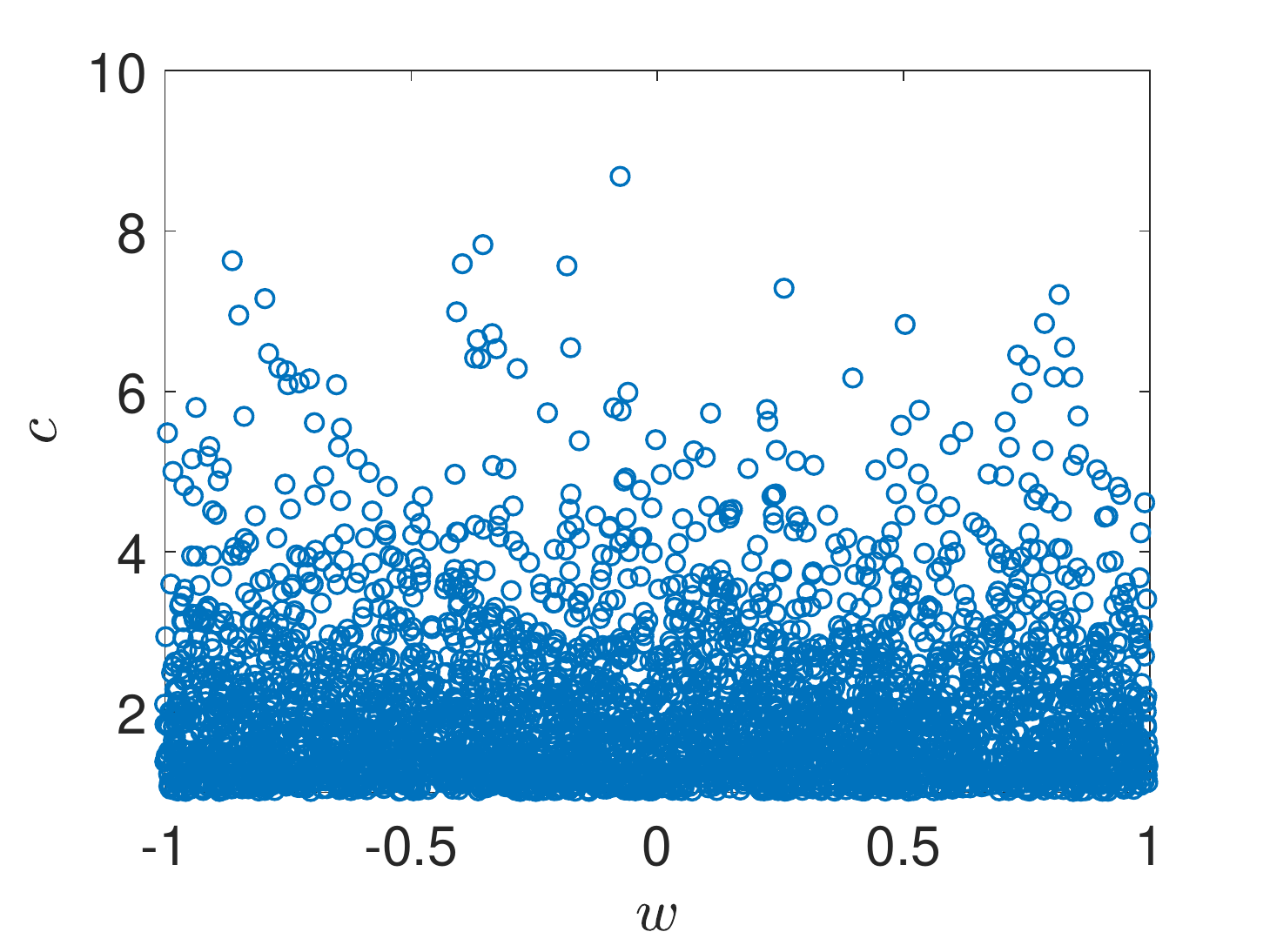}
\includegraphics[scale=0.5]{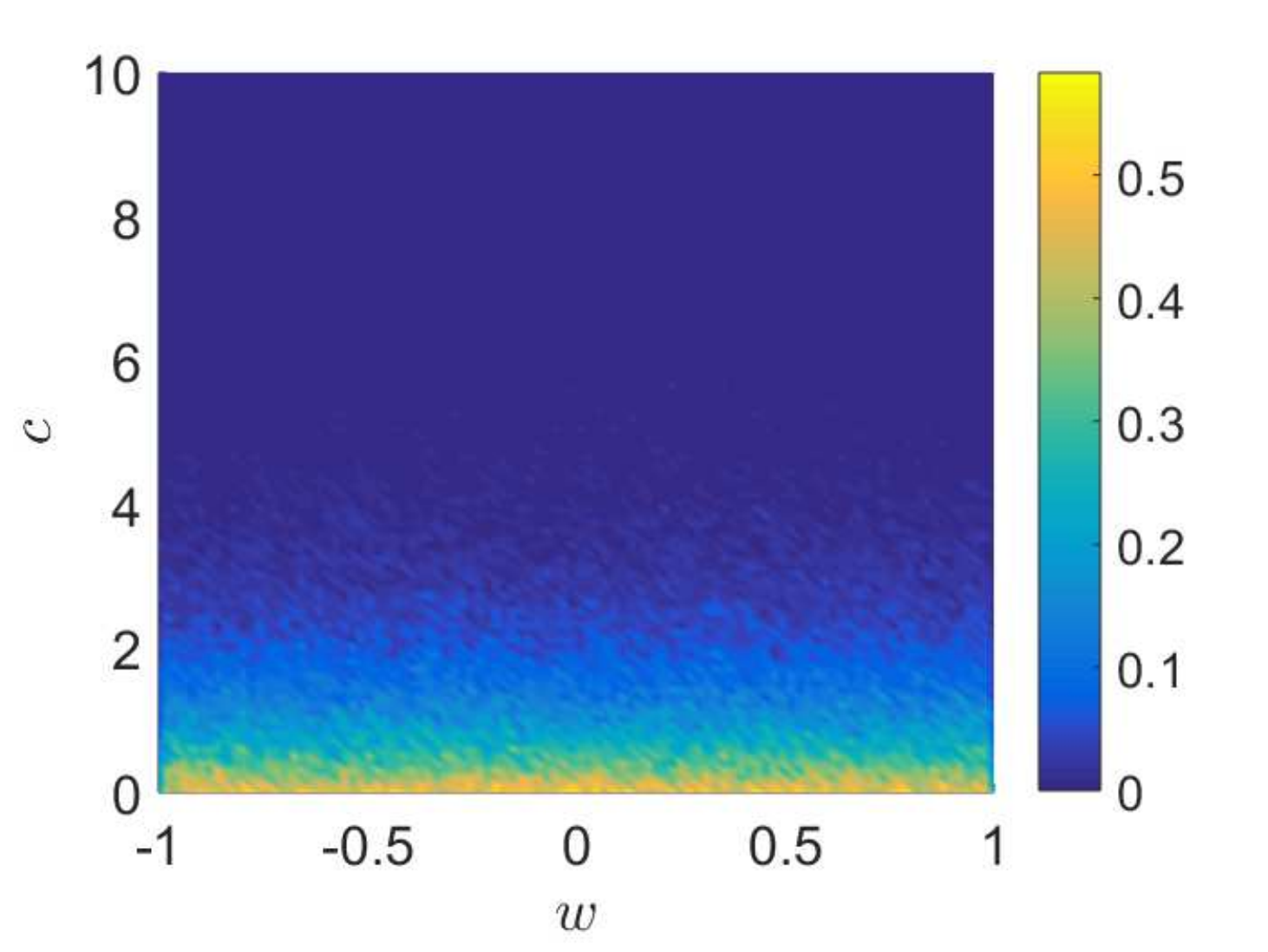}
\caption{\textbf{Test 1, Section~\ref{sect:test1}}. Initial distributions~\eqref{eq:p0.g1} (top row) and~\eqref{eq:p0.g2} (bottom row). Left column: $10^3$ particles sampled from the prescribed initial distributions. Right column: continuous numerical approximations of such distributions.}
\label{fig:p0.indep}
\end{figure}

As initial opinion distribution we take
$$ f_0(w):=f(0,\,w)=\frac{1}{2}\mathbb{1}_{[-1,\,1]}(w), $$
i.e., the uniform distribution in the interval $[-1,\,1]$ with mean $m=0$, which, as discussed in Section~\ref{sect:Boltzmann}, is conserved in time. The initial bivariate distribution function $p_0(w,\,c):=p(0,\,w,\,c)$ of the kinetic model is then either
\begin{equation}
    p_0(w,\,c)=\frac{1}{2}\mathbb{1}_{[-1,\,1]}(w)\frac{e^{-\frac{1}{c}}}{c^3},
    \label{eq:p0.g1}
\end{equation}
if the network connectivity is described by the distribution $g_1$ in~\eqref{eq:g12}, or
\begin{equation}
    p_0(w,\,c)=\frac{1}{2}\mathbb{1}_{[-1,\,1]}(w)e^{-c},
    \label{eq:p0.g2}
\end{equation}
if it is described by the distribution $g_2$ in~\eqref{eq:g12}. See Figure~\ref{fig:p0.indep}.

\begin{figure}[!t]
\includegraphics[scale=0.5]{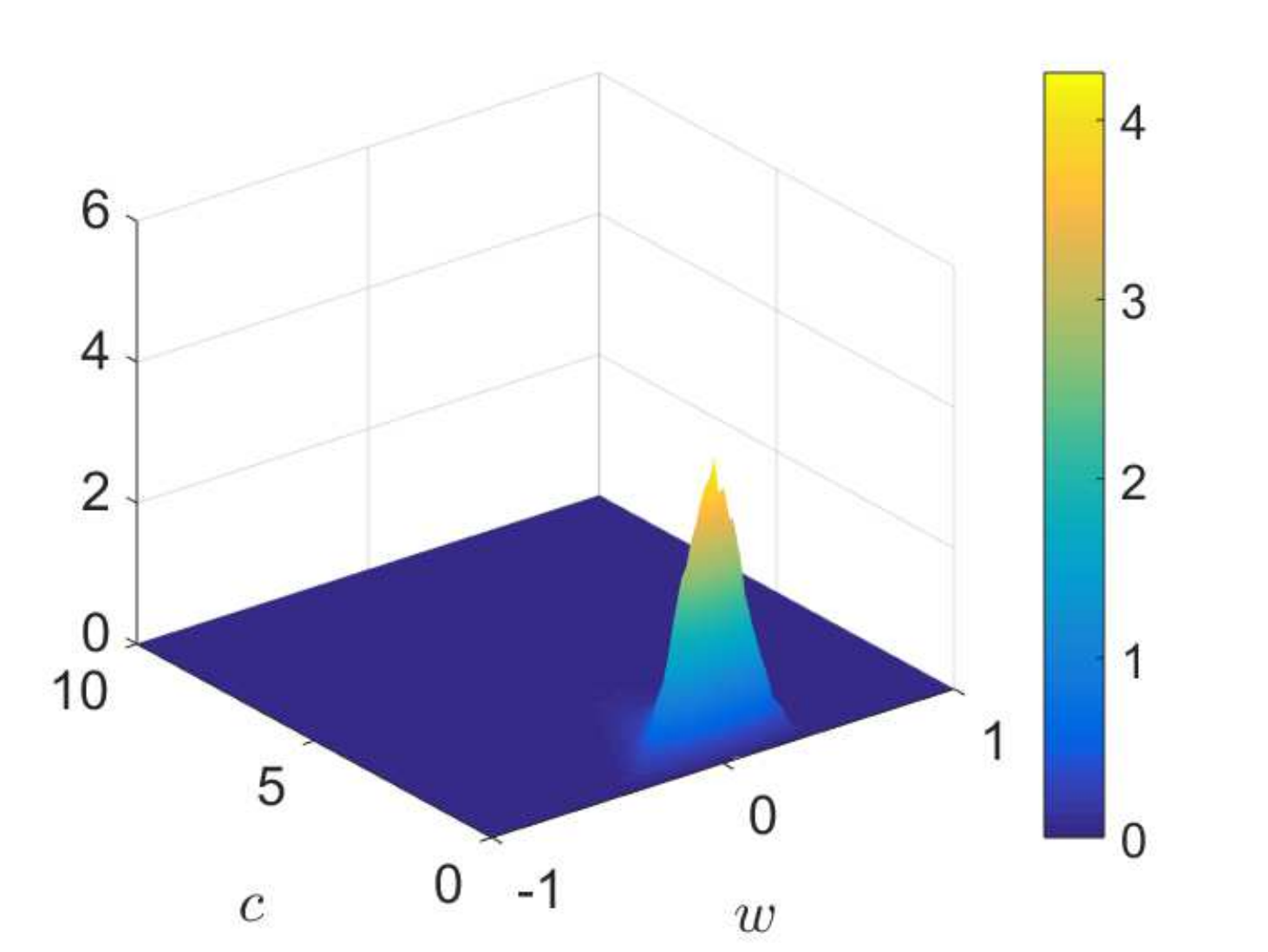}
\includegraphics[scale=0.5]{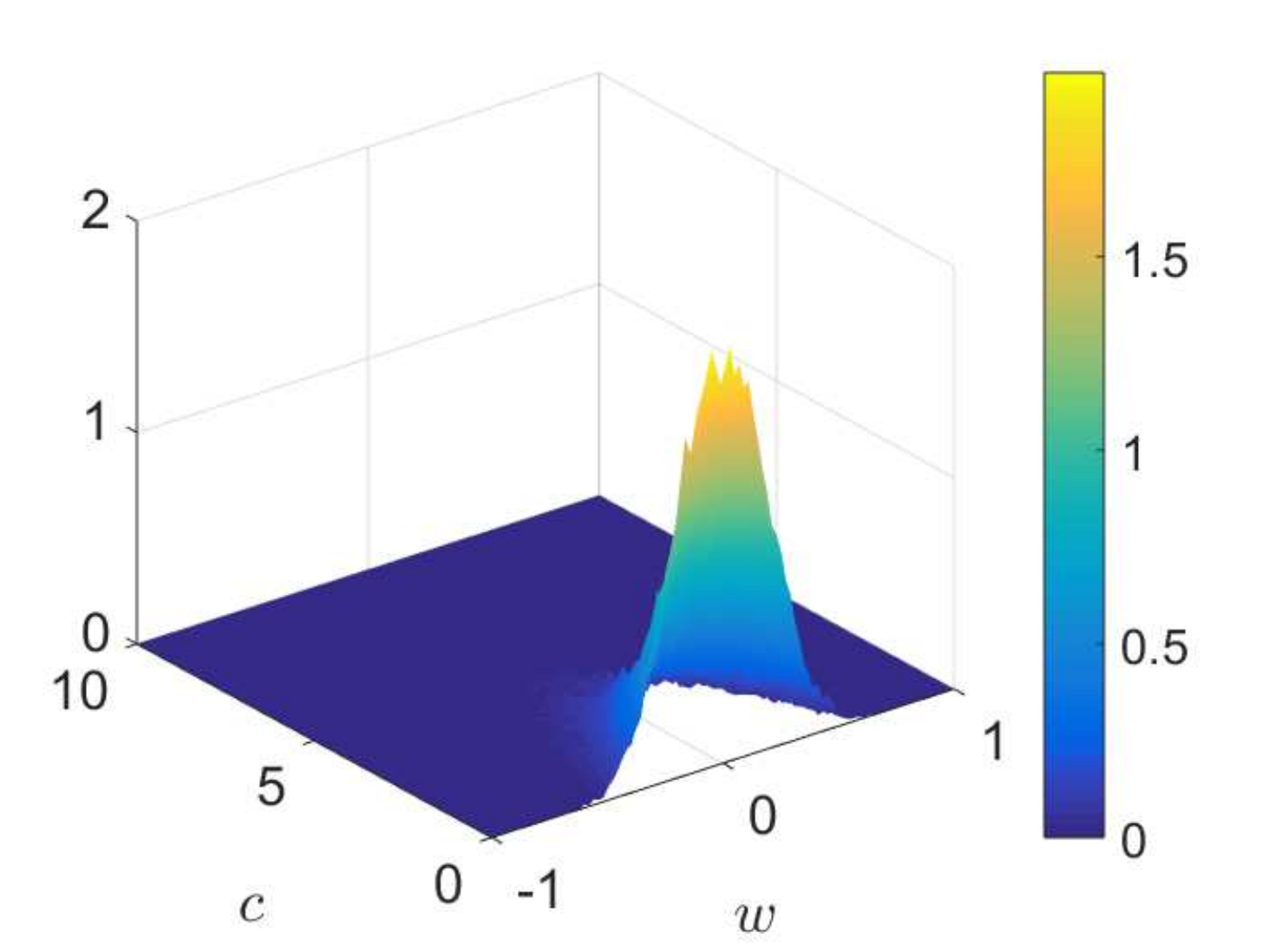}
\caption{\textbf{Test 1, Section~\ref{sect:test1}}. Large time bivariate solution $p^\infty(w,\,c)$ of the Boltzmann-type model~\eqref{eq:Boltzmann.1.strong}-\eqref{eq:Boltzmann.1.weak} with power law (left) and exponential (right) connectivity distribution, cf.~\eqref{eq:g12}, and with the independence assumption~\eqref{eq:p} of the variables $w$, $c$. The initial conditions are the distributions~\eqref{eq:p0.g1},~\eqref{eq:p0.g2}, respectively, displayed in Figure~\ref{fig:p0.indep}.}
\label{fig:pinf.indep}
\end{figure}

\begin{figure}[!t]
\includegraphics[scale=0.5]{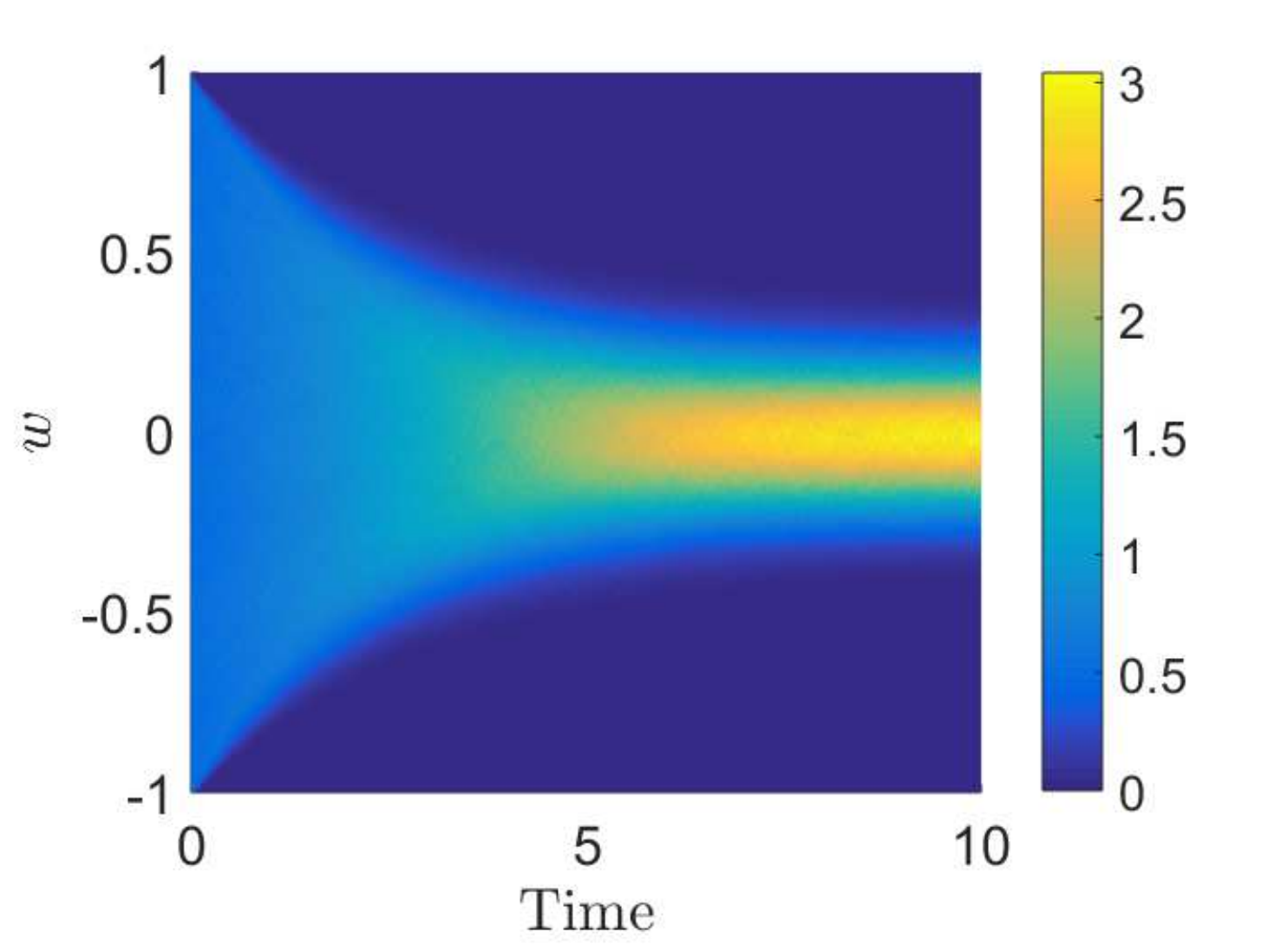}
\includegraphics[scale=0.5]{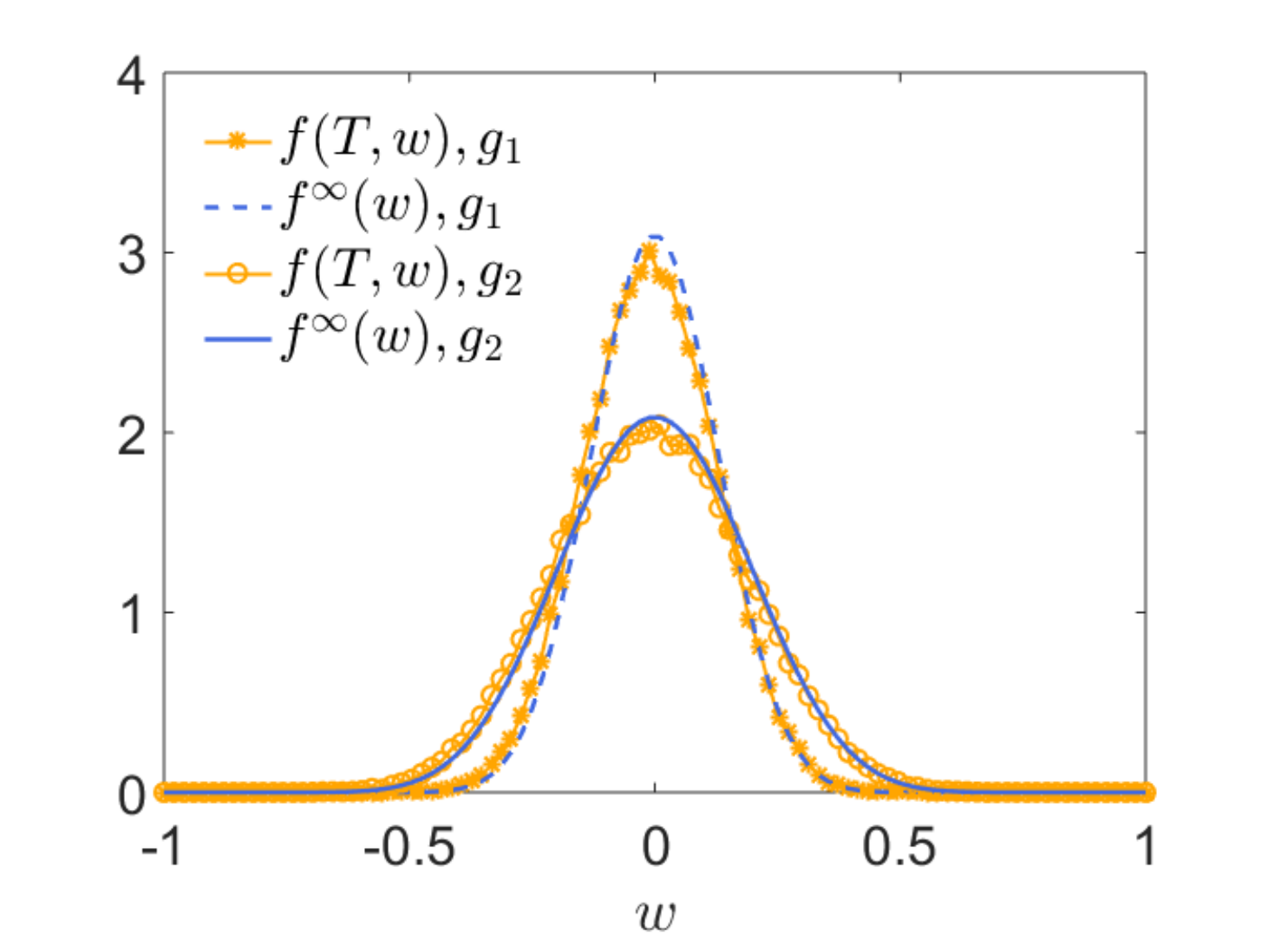}
\caption{\textbf{Test 1, Section~\ref{sect:test1}}. Left: transient trend of the marginal distribution $f(t,\,w)$ in the time interval $[0,\,10]$ with power law connectivity distribution. Right: large time opinion distribution (at the final computational time $T=10$) in the two cases of connectivity distribution ($g_1$: power law; $g_2$: exponential). The blue curves are the analytical steady distribution plotted from~\eqref{eq:finf.beta.1}.}
\label{fig:test1_transient}
\end{figure}

Figure~\ref{fig:pinf.indep} shows an approximation, for a sufficiently large time, of the asymptotic bivariate distribution $p^\infty(w,\,c)=f^\infty(w)g_{1,2}(c)$, where $f^\infty(w)$ is obtained by solving numerically the scaled Boltzmann equation~\eqref{eq:Boltzmann.3} with $\gamma=10^{-3}$. Not surprisingly, it turns out to be close to the asymptotic distribution~\eqref{eq:finf.beta.1} computed in the quasi-invariant opinion regime from the Fokker-Planck equation~\eqref{eq:FP.1}, as it is further clearly shown by Figure~\ref{fig:test1_transient}. The latter displays the transient behavior of $f(t,\,w)$ in the time interval $[0,\,10]$ and the asymptotic profile of $f^\infty(w)$, which agrees perfectly with~\eqref{eq:finf.beta.1} computed using the values of $\cC$, $\cB$ evaluated from~\eqref{eq:C},~\eqref{eq:B} by numerical integration of the functions $\kappa$, $\beta$, $g_{1,2}$ of the present test and moreover with $\lambda=1$.

It is interesting to observe, from the right panel of Figure~\ref{fig:test1_transient}, that, thanks to its heavier tail, the power law distribution $g_1$ induces a greater opinion consensus around the conserved mean $m=0$ than that produced by the exponential distribution $g_2$. This is somehow reminiscent of the dichotomy between clustering and consensus dynamics analyzed in~\cite{motsch2014SIREV}.

\subsection{Test 2: Opinion dynamics with dependent \texorpdfstring{$\boldsymbol{w}$}{}, \texorpdfstring{$\boldsymbol{c}$}{}}
\label{sect:test2}
\begin{figure}[!t]
\includegraphics[scale=0.5]{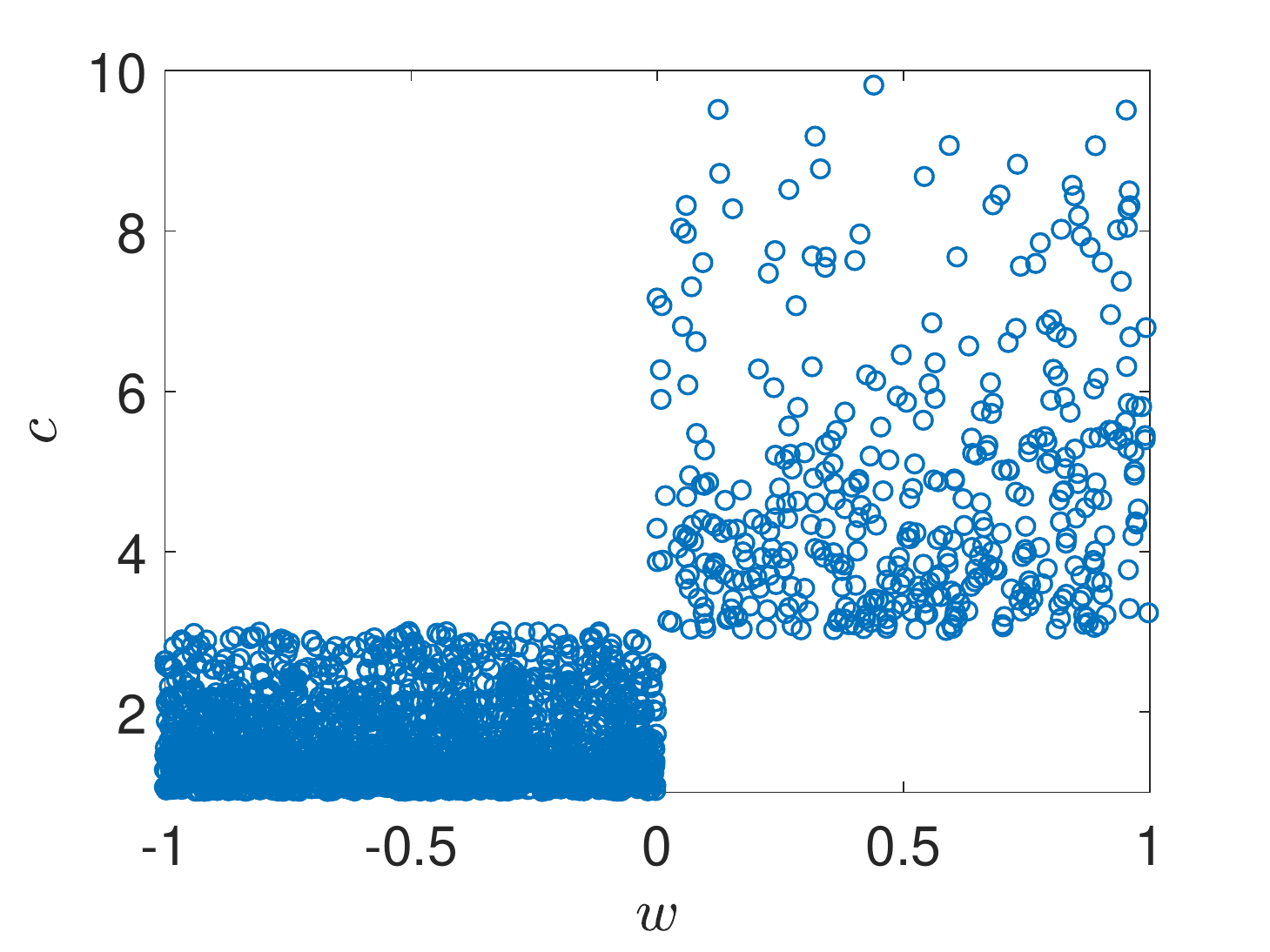}
\includegraphics[scale=0.5]{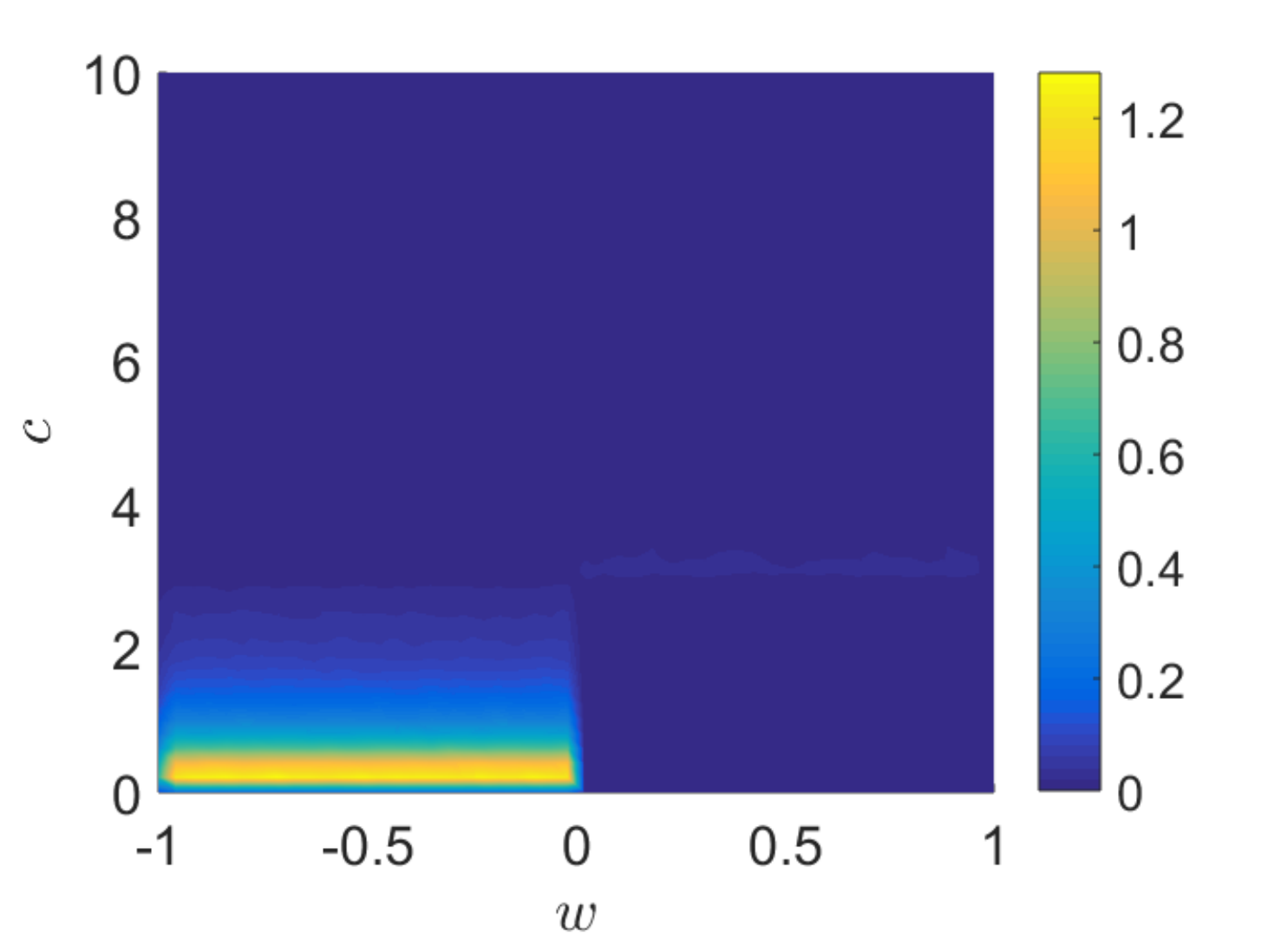} \\
\includegraphics[scale=0.5]{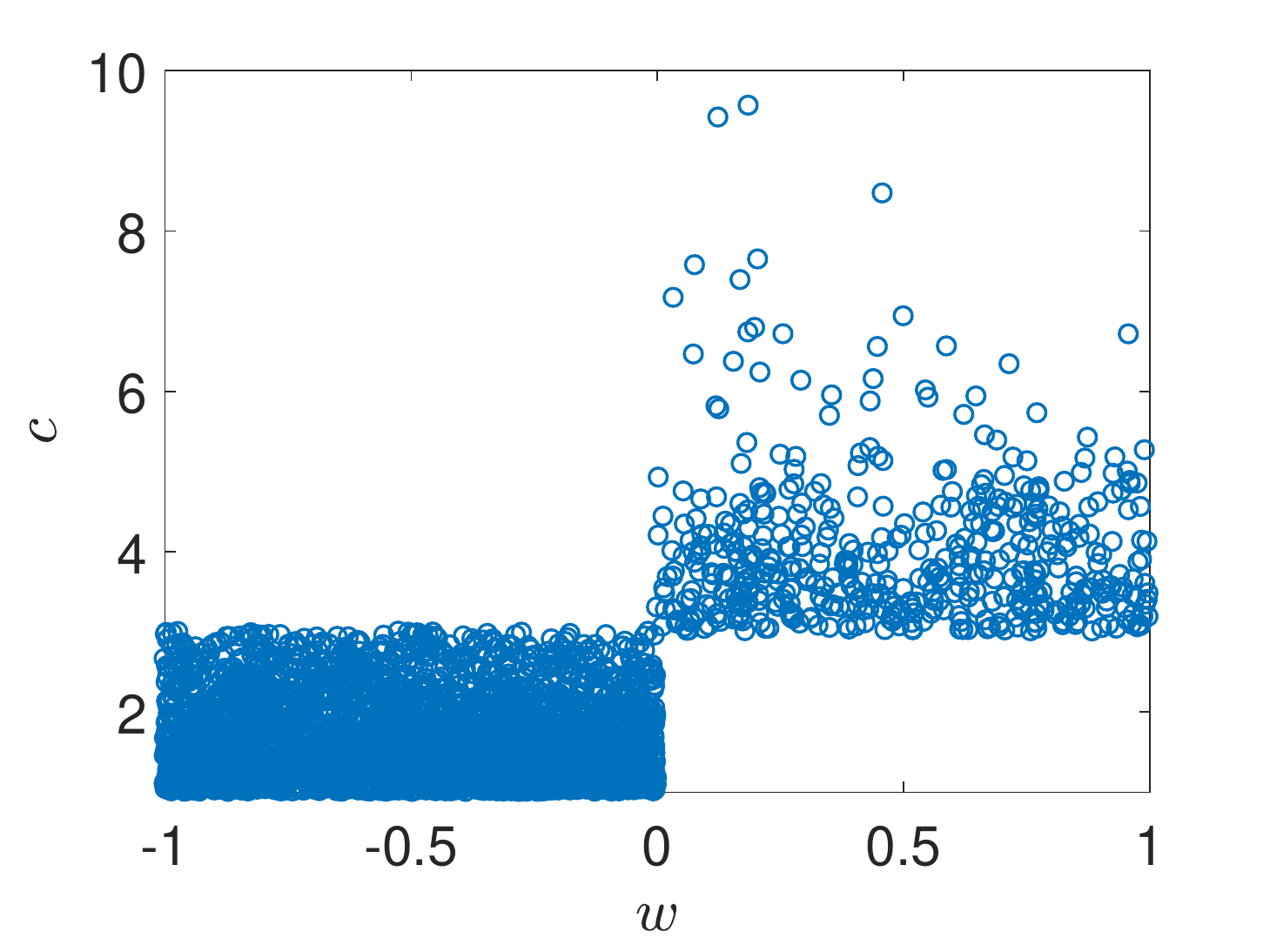}
\includegraphics[scale=0.5]{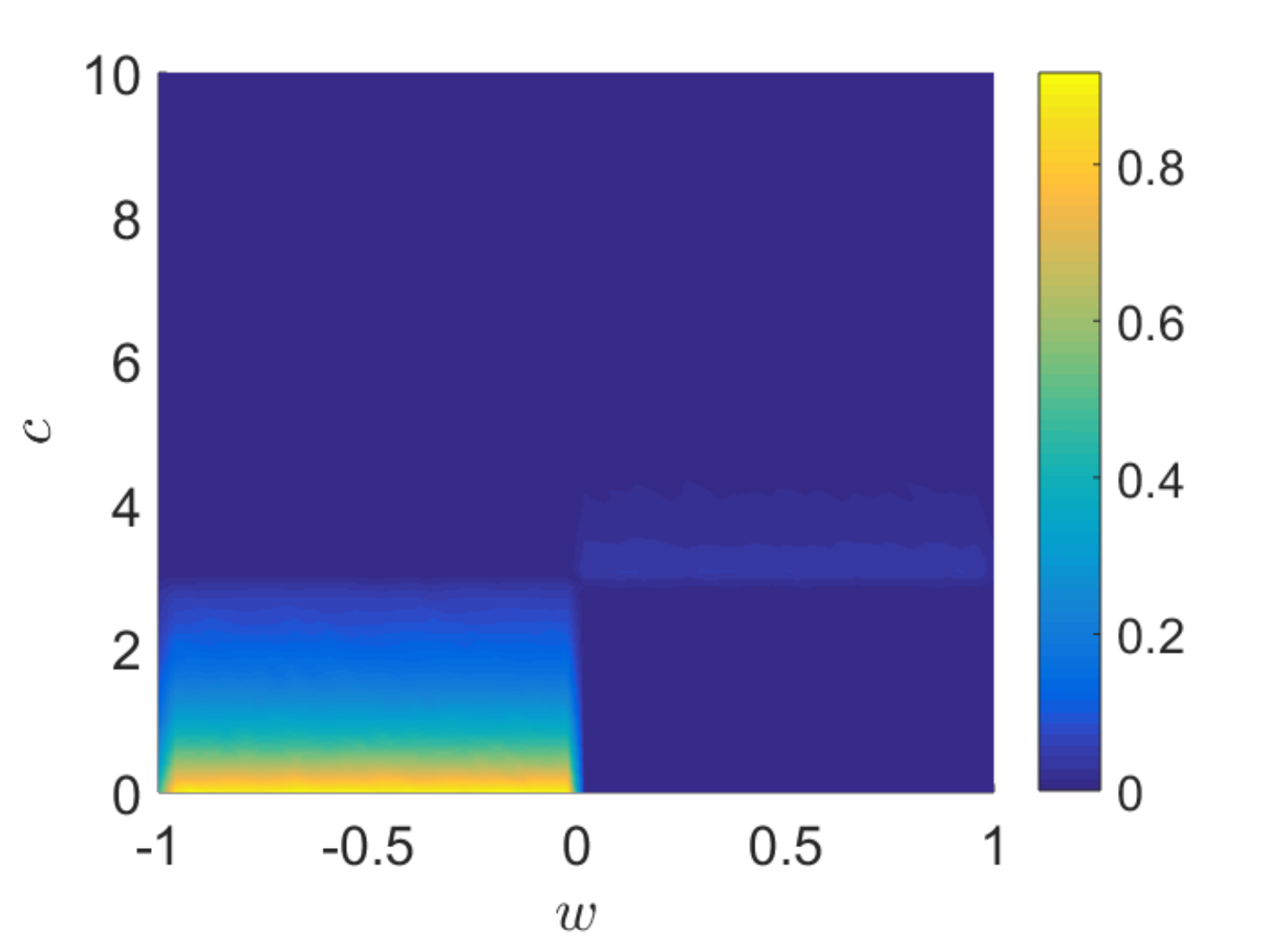}
\caption{\textbf{Test 2, Section~\ref{sect:test2}}. Initial distributions~\eqref{eq:p0.dep} with $c_0=3$ and connectivity distribution given by either the power law $g_1$ in~\eqref{eq:g12} (top row) or the exponential law $g_2$ in~\eqref{eq:g12} (bottom row). Left column: $10^3$ particles sampled from the prescribed initial distributions. Right column: continuous numerical approximations of such distributions.}
\label{fig:p0.dep}
\end{figure}

In this second test we consider the Boltzmann-type model~\eqref{eq:binary.w.1}-\eqref{eq:Boltzmann.1.strong} for the bivariate kinetic distribution $p(t,\,w,\,c)$ \textit{without} the independence assumption~\eqref{eq:p} between the variables $w$ and $c$. The reference theoretical discussion is the one set forth in Section~\ref{sect:p.gen}. In particular, we choose as initial condition a distribution $p_0(w,\,c):=p(0,\,w,\,c)$ which does not write as the product of its marginals, precisely:
\begin{equation}
    p_0(w,\,c)=\left[\mathbb{1}_{[-1,\,0]\times [0,\,c_0)}(w,\,c)+\mathbb{1}_{(0,\,1]\times [c_0,\,+\infty)}(w,\,c)\right]g_{1,2}(c),
    \label{eq:p0.dep}
\end{equation}
where $g_{1,2}(c)$ is either connectivity distribution given in~\eqref{eq:g12}. The function
$$ f_{c,0}(w)=\mathbb{1}_{[-1,\,0]\times [0,\,c_0)}(w,\,c)+\mathbb{1}_{(0,\,1]\times [c_0,\,+\infty)}(w,\,c), $$
with $c_0\in\R_+$, is the initial conditional distribution of $w$ given $c$. In practice, such an $f_{c,0}$ says that individuals with a connectivity lower than $c_0$ have initially an opinion uniformly distributed in the interval $[-1,\,0]$ whereas individuals with a connectivity greater than or equal to $c_0$ have initially an opinion uniformly distributed in the interval $[0,\,1]$. This models polarized opinions around the mean values $\mp\frac{1}{2}$ of $f_{c,0}$ (for $c<c_0$ and $c\geq c_0$, respectively) depending on the connectivity distribution.

Figure~\ref{fig:p0.dep} shows either distribution~\eqref{eq:p0.dep} with $c_0=3$ for the two different choices~\eqref{eq:g12} of the distribution $g$. It is worth pointing out that with this threshold $c_0$ the initial fraction of individuals with opinion $w\leq 0$, namely $\int_{-1}^0\int_{\R_+}p_0(w,\,c)\,dc\,dw$, is invariably larger than that of individuals with opinion $w>0$, namely $\int_0^1\int_{\R_+}p_0(w,\,c)\,dc\,dw$, for both choices of the distribution $g$.

We compute the large time trend of the system choosing the connectivity-based interaction function $\kappa$~\eqref{eq:k} and the local diffusion coefficient $D_\op$~\eqref{eq:D} with $\beta$ given by~\eqref{eq:test1.beta}. Notice in particular that the choice~\eqref{eq:k} of $\kappa$ makes this numerical experiment depart from the analysis performed in Section~\ref{sect:p.gen}, where the function $\kappa$ was supposed to depend only on the connectivity of one individual of the interacting pair. This gives us the opportunity to explore a scenario which was not possible to cover analytically in detail.

\begin{figure}[!t]
\includegraphics[scale=0.5]{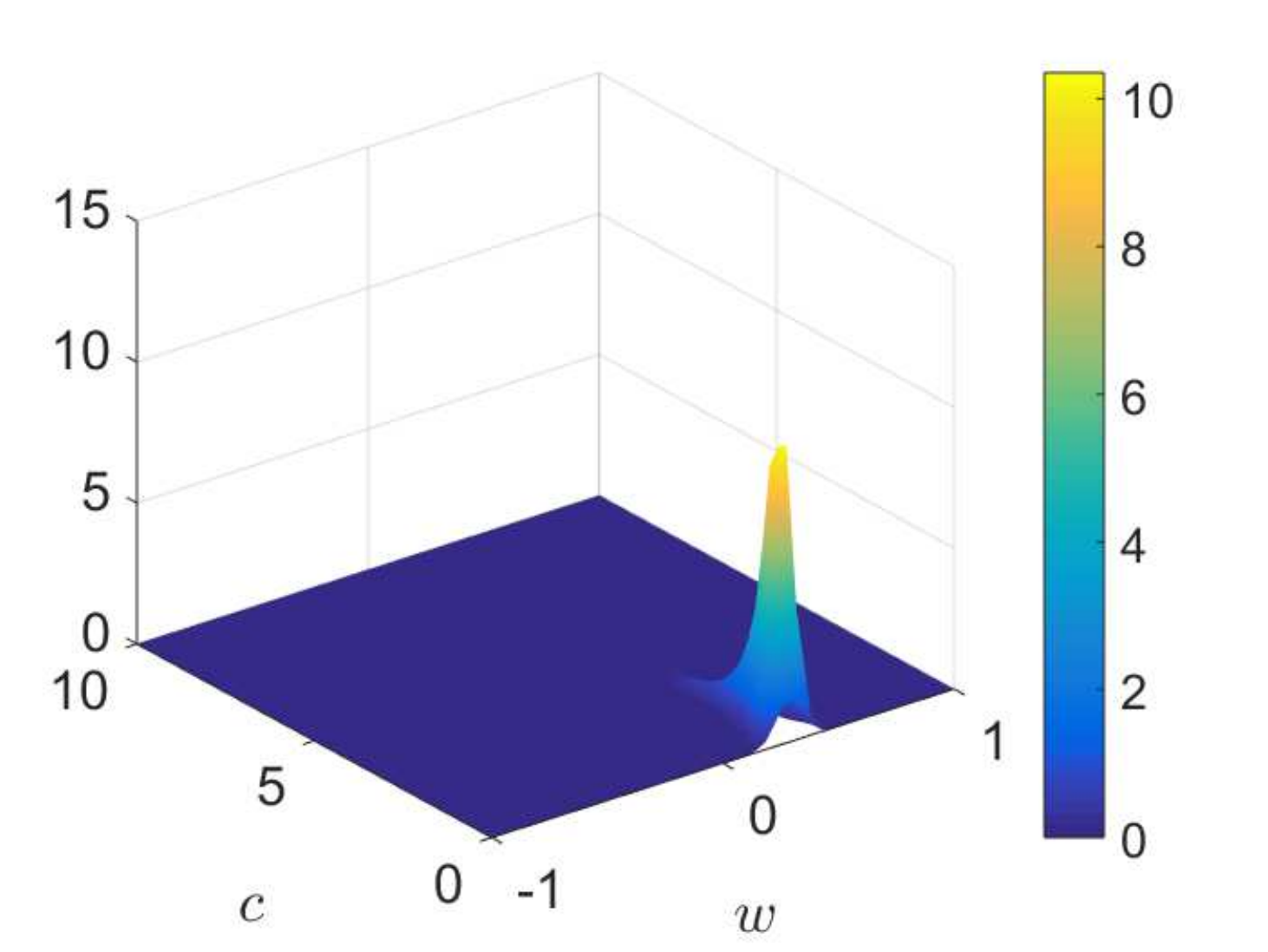}
\includegraphics[scale=0.5]{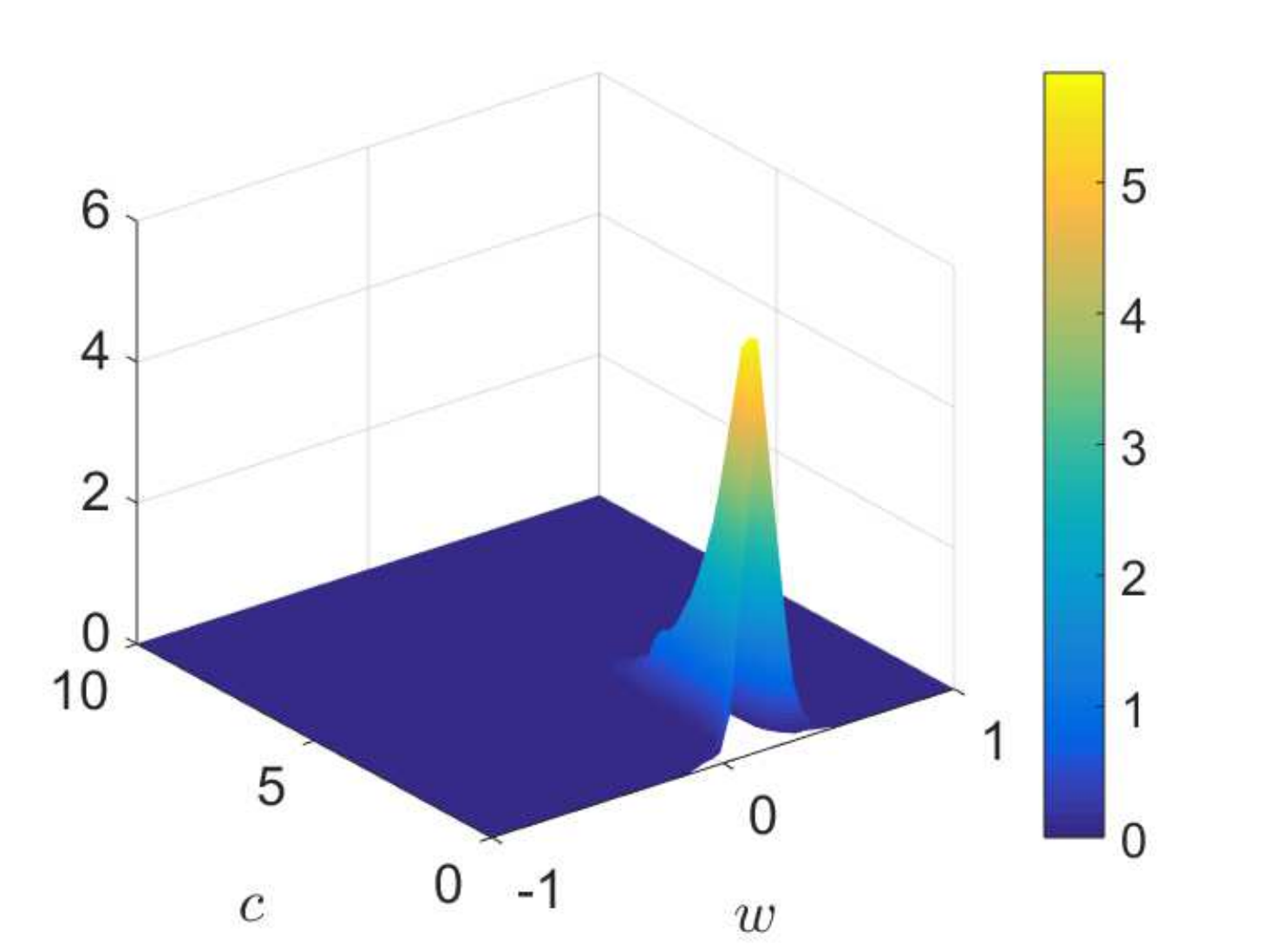}
\caption{\textbf{Test 2, Section~\ref{sect:test2}}. Large time bivariate solution $p^\infty(w,\,c)$ of the Boltzmann-type model~\eqref{eq:Boltzmann.1.strong}-\eqref{eq:Boltzmann.1.weak} with power law (left) and exponential (right) connectivity distribution, cf.~\eqref{eq:g12} and without the independence assumption~\eqref{eq:p} of the variables $w$, $c$. The initial conditions are the distributions~\eqref{eq:p0.dep} displayed in Figure~\ref{fig:p0.dep}.}
\label{fig:pinf.dep}
\end{figure}

\begin{figure}[!t]
\includegraphics[scale=0.5]{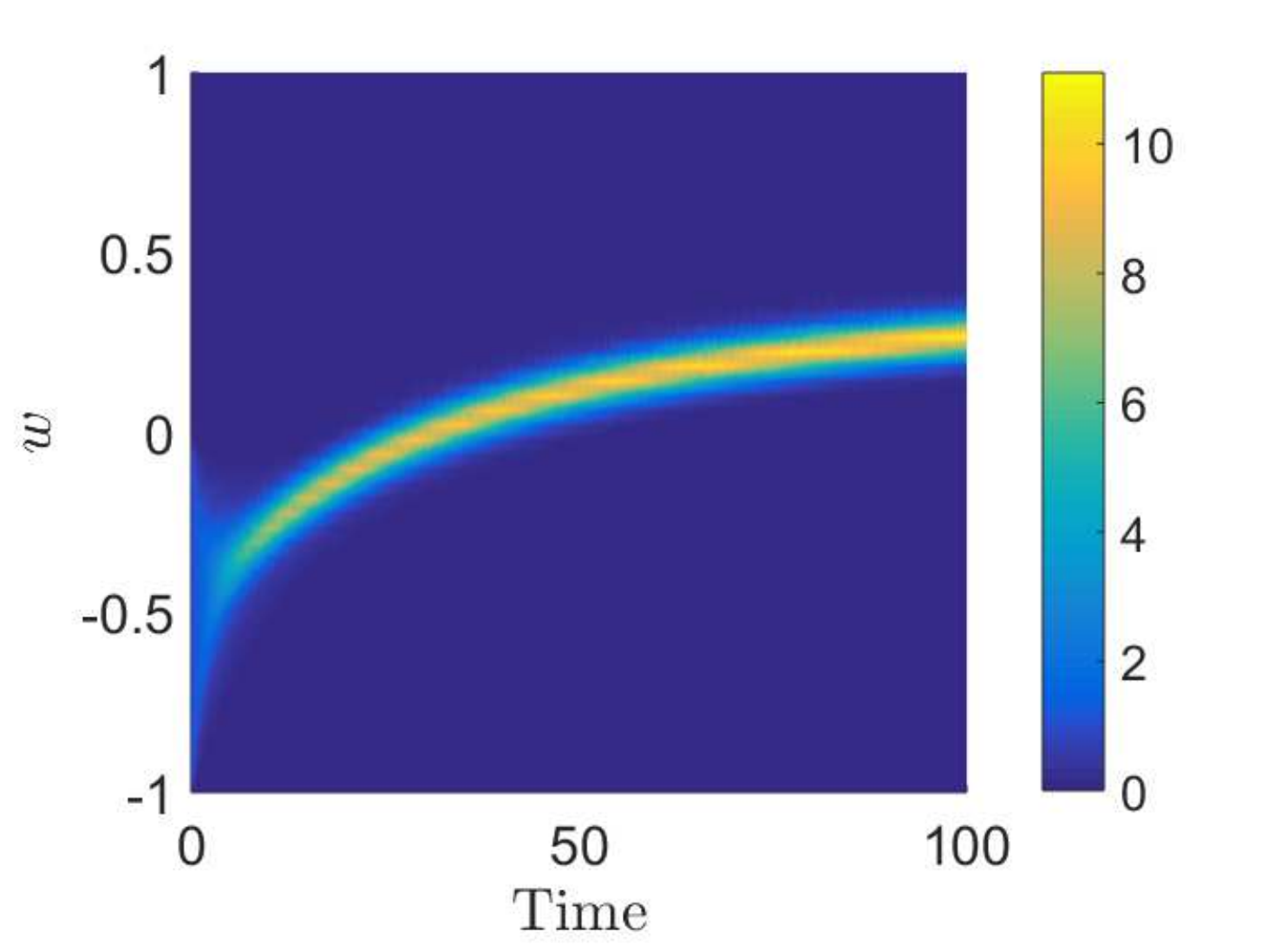}
\includegraphics[scale=0.5]{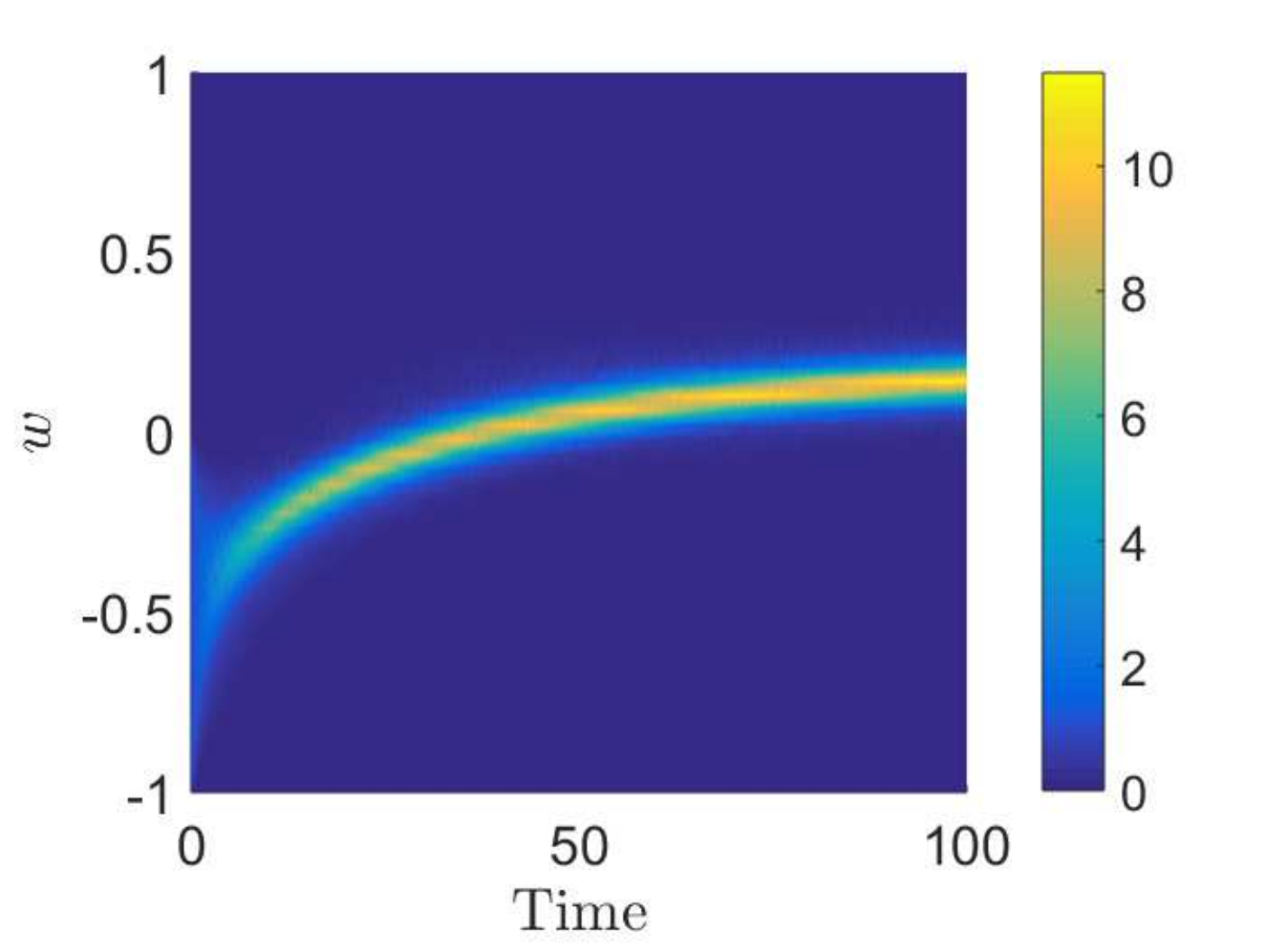}
\caption{\textbf{Test 2, Section~\ref{sect:test2}}. Time evolution of the marginal opinion density $f(t,\,w)$ in the time interval $[0,\,100]$ for power law (left) and exponential (right) connectivity distribution.}
\label{fig:ptrans.dep}
\end{figure}

\begin{figure}[!t]
\centering
\includegraphics[scale=0.5]{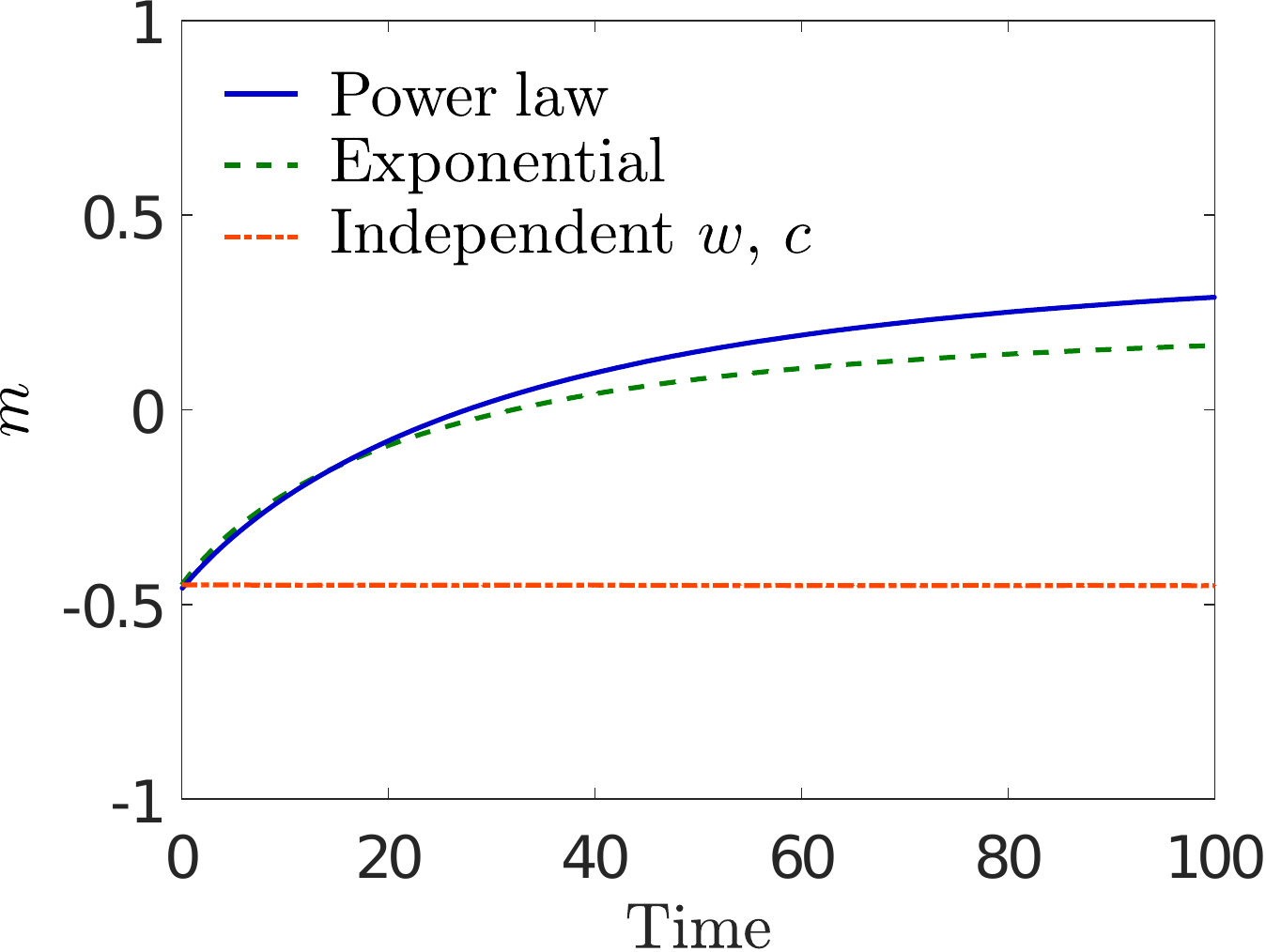}
\caption{\textbf{Test 2, Section~\ref{sect:test2}}. Time evolution of the mean opinion in the time interval $[0,\,100]$ with the two connectivity distributions given in~\eqref{eq:g12}. The evolution of the mean opinion in the case of independent $w$, $c$, cf. Section~\eqref{sect:Boltzmann}, is also reported for duly comparison.}
\label{fig:m.dep}
\end{figure}

Figure~\ref{fig:pinf.dep} shows the large time solution of the Boltzmann-type model~\eqref{eq:binary.w.1}-\eqref{eq:Boltzmann.1.strong} computed under the time scaling $\tau=\gamma t$ with $\gamma=10^{-3}$ (cf. the quasi-invariant opinion limit described in Section~\ref{sect:FP}). The numerical solution is shown at the computational time $T=100$, which, owing to the above scaling, yields a good approximation of the asymptotic profile. It is interesting to observe that, in contrast to the test considered in Section~\ref{sect:test1}, here the statistical dependence of the variables $w$, $c$ tends to drive the kinetic distribution $p$ toward the region of the state space where $w\geq 0$. This happens in spite of the fact that initially such a region contains the lower fraction of individuals, who are however the most influential ones of the social network, namely those with an over-threshold connectivity $c$. This trend is even more evident in Figure~\ref{fig:ptrans.dep}, which shows the time evolution of the marginal opinion distribution $f(t,\,w)$, and in Figure~\ref{fig:m.dep}, which shows the time evolution of the mean opinion $m(t)=\int_{-1}^1wf(t,\,w)\,dw$. Unlike the case addressed analytically and numerically in Sections~\ref{sect:Boltzmann},~\ref{sect:test1}, here $m$ is no longer constant in time (as expected) and, in particular, is attracted toward a nonnegative asymptotic value (with small variance, cf. Figure~\ref{fig:ptrans.dep}) dictated by the initial distribution of the influencers, i.e., the individuals with $c\geq c_0$.

It is also interesting to observe from Figure~\ref{fig:m.dep} that with the power law connectivity distribution $g_1$ the mean opinion $m$ tends asymptotically to a value closer to the initial mean opinion of the influencers, namely $\frac{1}{2}$, than with the exponential connectivity distribution $g_2$. This is consistent with the fact that the heavier tail of $g_1$ makes the presence of influencers, i.e., individuals with a very high connectivity, much more probable.

\subsection{Test 3: Spreading of the popularity}
\label{sect:test3}
In this test we consider model~\eqref{eq:system_Boltzmann} for the spreading of the popularity of a product in connection with the opinion dynamics investigated so far. In particular, in contrast to the analysis performed in Section~\ref{sect:popularity}, here we solve the dynamically coupled model, i.e., we do not assume that opinion dynamics are necessarily a much quicker process than the spreading of the popularity. This allows us to account for a richer time trend of the relevant statistical quantities than the simple exponential decay found in Section~\ref{sect:Boltzmann.v}.

We adopt the power law connectivity distribution $g_1$ given in~\eqref{eq:g12} and, as far as the opinion distribution is concerned, we take as initial condition the corresponding probability density function~\eqref{eq:p0.dep} with $c_0=3$, see also the top row of Figure~\ref{fig:p0.dep}. This means, in particular, that we consider the case of dependent variables $w$, $c$. In parallel, we assume that initially the popularity $v$ is uniformly distributed in the interval $[0,\,10]$, hence
$$ h_0(v):=h(0,\,v)=\frac{1}{10}\mathbb{1}_{[0,\,10]}(v). $$
We fix the other parameters of the microscopic interactions, cf.~\eqref{eq:binary.v}-\eqref{eq:P}, as follows: $\epsilon=10^{-3}$ (the scaling parameter of the quasi-invariant regime, cf. Sections~\ref{sect:Boltzmann.v},~\ref{sect:FP.v}), $\mu=10^{-5}$, whence $\mu_0=\mu/\epsilon=10^{-2}$, $\nu=10^{-3}$, whence $\nu_0=\nu/\epsilon=1$, $\Delta=0.15$, $\varsigma^2=10^{-5}$, whence $\zeta=\varsigma^2/\epsilon=10^{-2}$. Finally, we choose the diffusion coefficient $D_\pop(v)=v$. Notice that this choice of the microscopic parameters implies in particular $\mu_0/\zeta=1$, which is the condition found in Section~\ref{sect:FP.v} for obtaining an asymptotic distribution of the popularity with Pareto exponent $3$, the value supported by empirical observations.

\begin{figure}[!t]
\centering
\subfigure[$\hat{w}=-\frac{1}{2}$]{\includegraphics[scale=0.45]{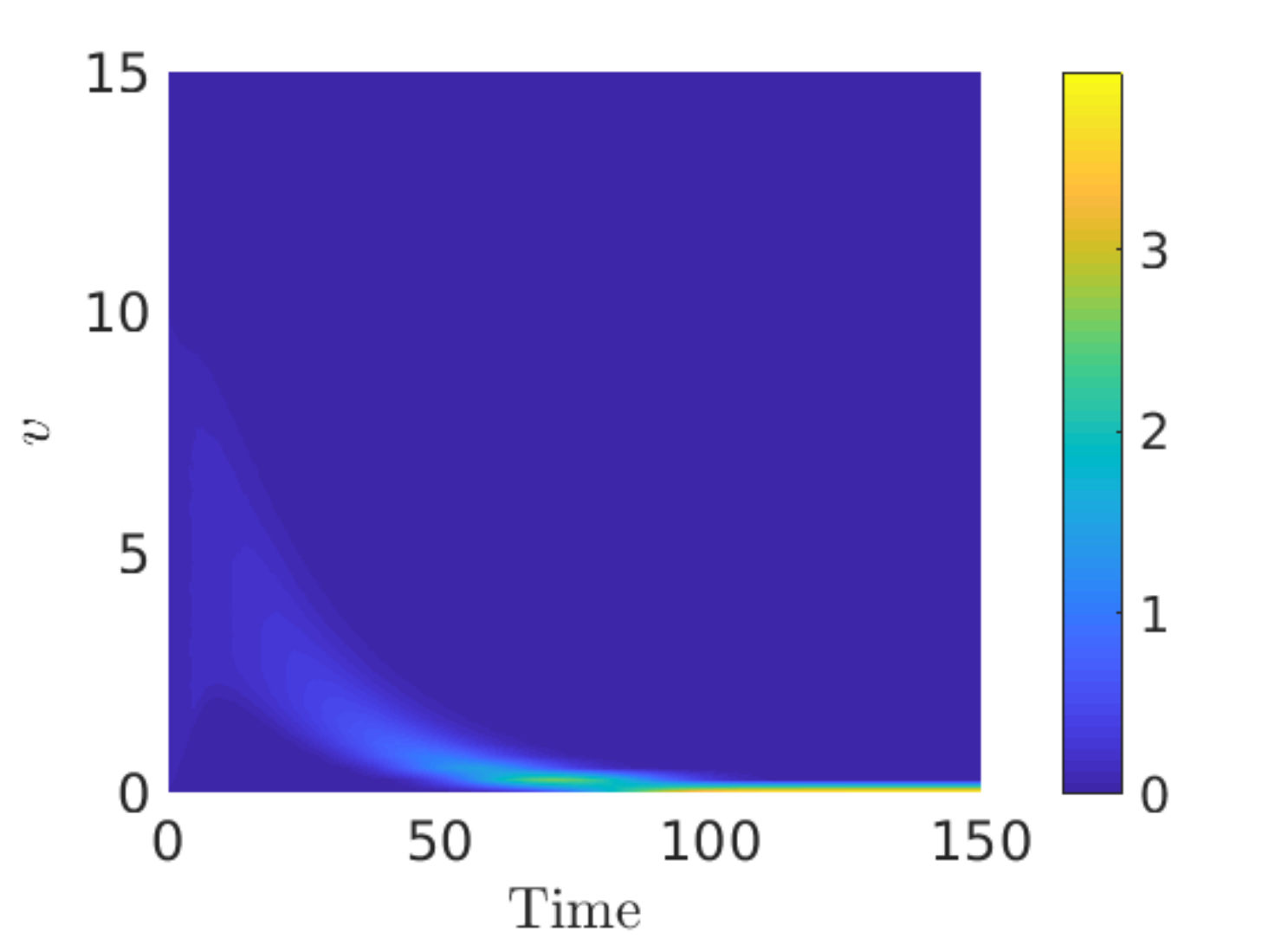}}
\subfigure[$\hat{w}=0$]{\includegraphics[scale=0.45]{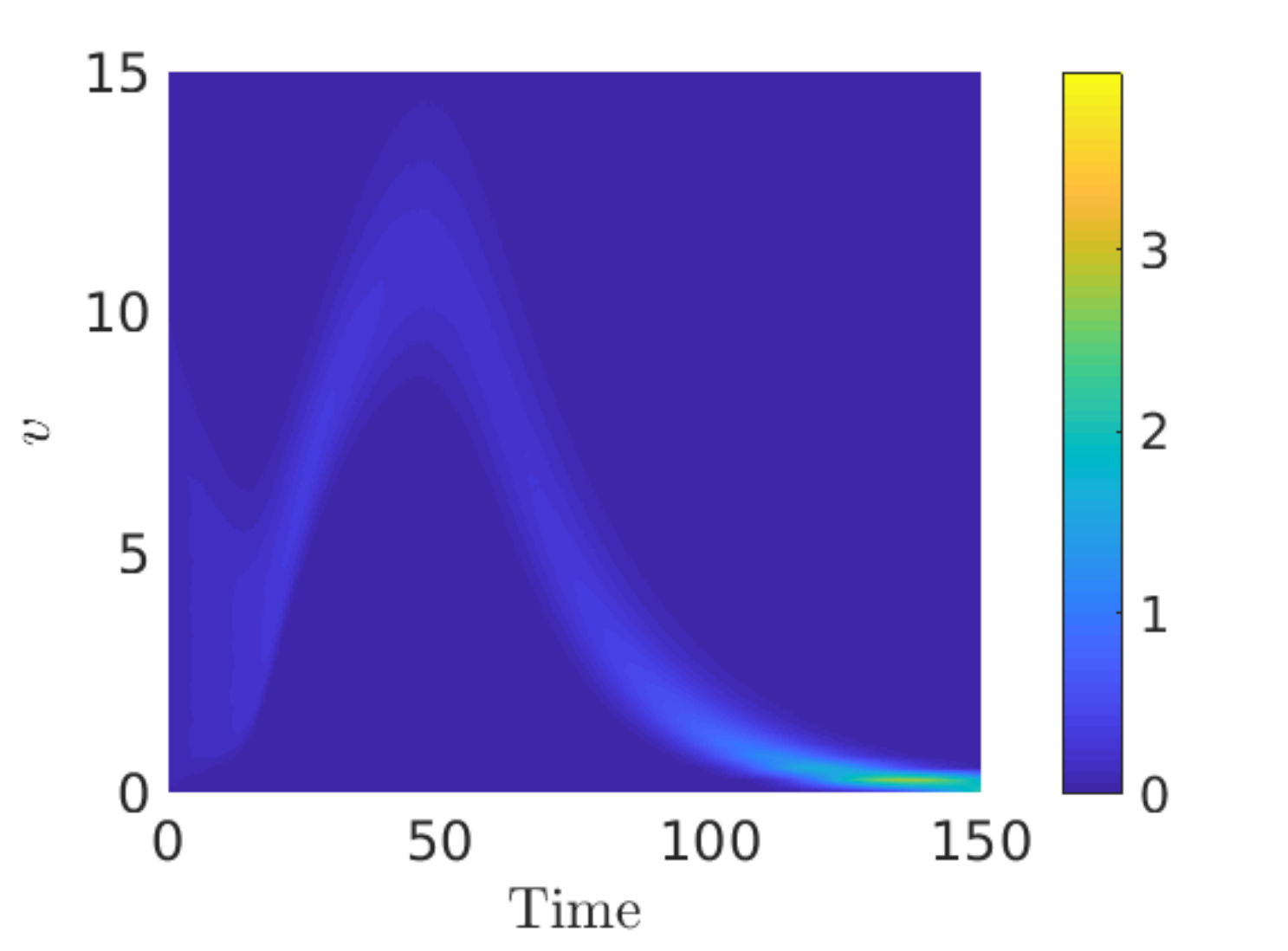}} \\
\subfigure[$\hat{w}=\frac{1}{2}$]{\includegraphics[scale=0.45]{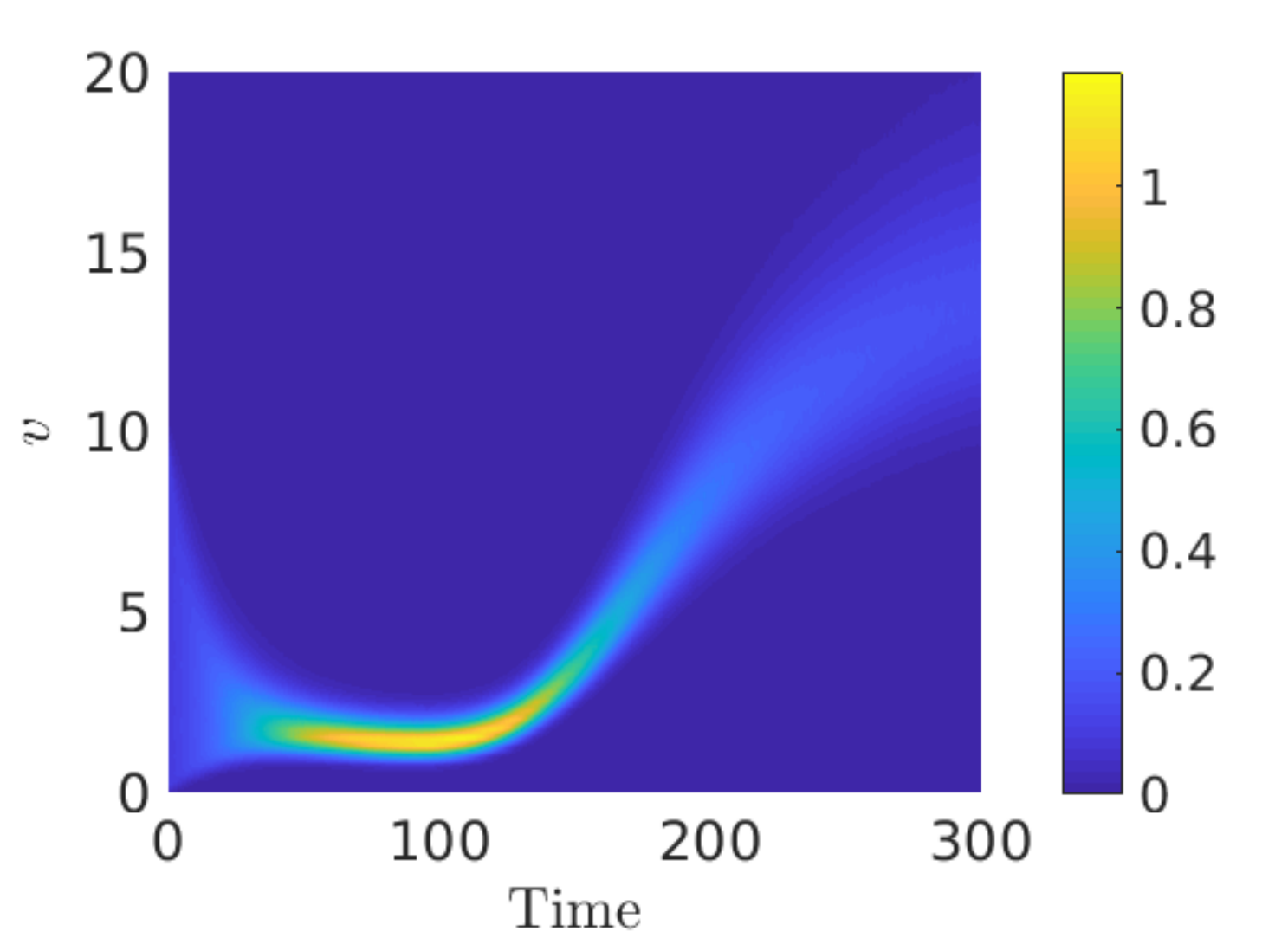}}
\subfigure[$h^{\infty}$]{\includegraphics[scale=0.45]{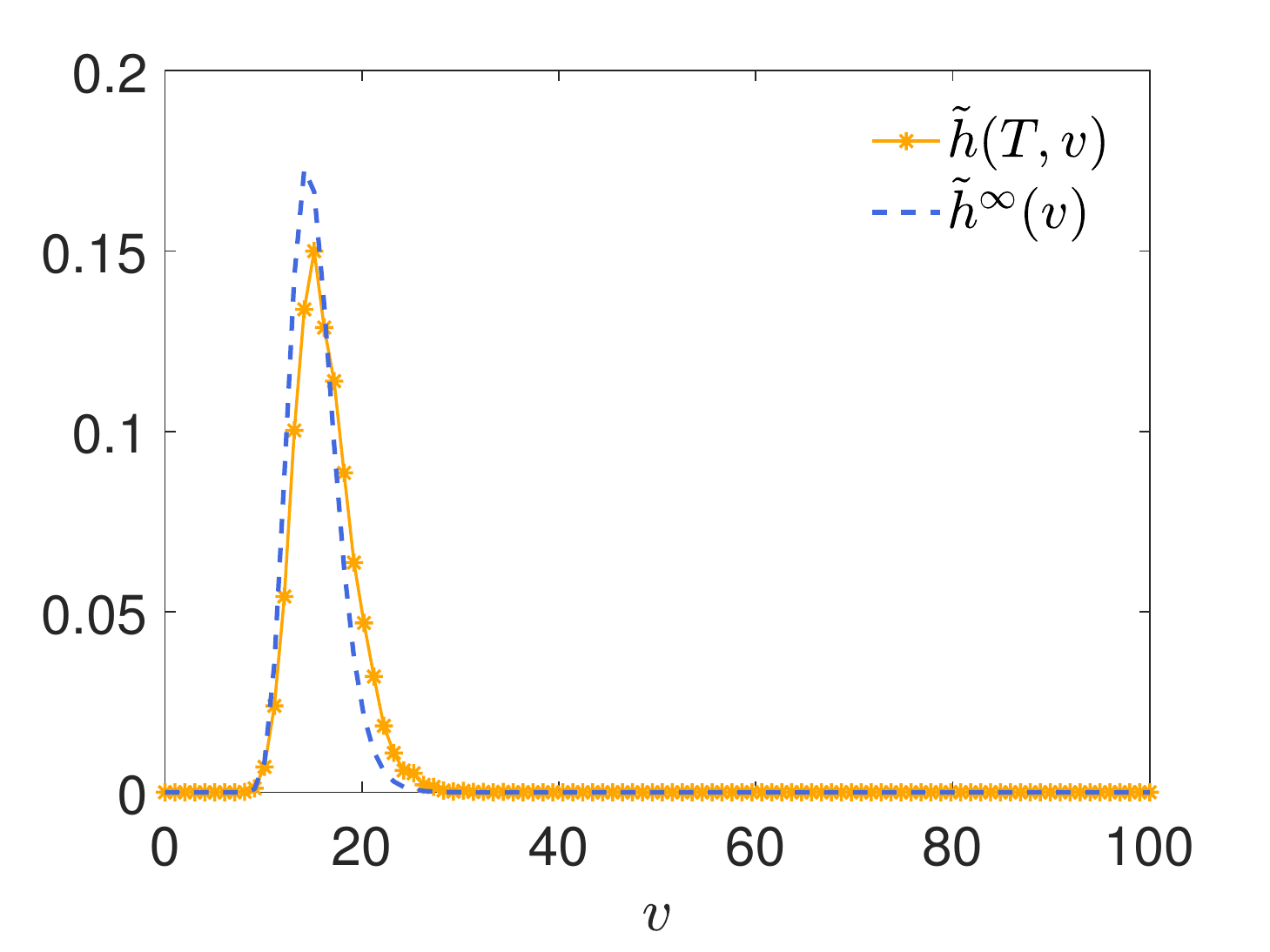}}
\caption{\textbf{Test 3, Section~\ref{sect:test3}}. (a)--(c) Time evolution of the popularity distribution $h(t,\,v)$ coupled to the opinion dynamics examined in Test 2, Section~\ref{sect:test2} for the three different choices of the target opinion $\hat{w}=-\frac{1}{2},\,0,\,\frac{1}{2}$. (d) Analytical and numerical asymptotic distributions of the popularity computed respectively from~\eqref{eq:h.Pareto} and by numerical integration of model~\eqref{eq:system_Boltzmann} up to the computational time $T=700$.}
\label{fig:pop}
\end{figure}

The initial condition $p_0(w,\,c)$ in~\eqref{eq:p0.dep} with the fat-tailed connectivity distribution $g_1(c)$, cf.~\eqref{eq:g12}, implies that initially the mean opinion of the population is
$$ m_0:=\int_{\R_+}\int_{-1}^1wp_0(w,\,c)\,dw\,dc=\frac{1}{2}-\Gamma\left(2,\,\frac{1}{3}\right)\approx -0.45, $$
where $\Gamma(s,\,x):=\int_x^{+\infty}t^{s-1}e^{-t}\,dt$ is the incomplete gamma function. By looking only at this aggregate characteristic of the society one may argue that a good strategy for enhancing the popularity of a product is to target an opinion close to $m_0$. For instance, one may take $\hat{w}=-\frac{1}{2}$ in~\eqref{eq:P}. Nevertheless, our mathematical model shows that this is by no means a good choice, in fact in such a case the popularity decays rapidly to zero, cf. Figure~\ref{fig:pop}a.

Another possibly common strategy may be to target the neutral opinion $\hat{w}=0$ in the hope that this may render the product more generically suited to most individuals. Our model clearly shows that also this choice turns out to be ineffective, because, after a temporary rise, the popularity of the product invariably vanishes in the long run, cf. Figure~\ref{fig:pop}b.

The drawback of both strategies is that they choose the target opinion $\hat{w}$ by basically ignoring statistical facts related to the connectivity distribution in the population. Considering that most individuals are actually not influencers, in fact the fraction of agents with low connectivity ($c<c_0=3$) is
$$ \int_0^3\int_{-1}^1p_0(w,\,c)\,dw\,dc=\int_0^3g_1(c)\,dc=\Gamma\left(2,\,\frac{1}{3}\right)\approx 95\%, $$
one understands that the value $m_0$ is mainly determined by people who will not be able to dictate the opinion trends. Owing to this, the strategy of targeting the initial mean opinion of the influencers ($c\geq c_0$), namely $\hat{w}=\frac{1}{2}$, turns out to be successful in spite of the very small fraction of influencers in the social network (about $5\%$). In fact, as Figure~\ref{fig:pop}c shows, the popularity suffers from an initial short decrease due to the fact that the target opinion is somehow elitist at the beginning; then it goes through a non-zero plateau during the influencer-driven phase of reorganization of the opinions; and finally it steadily rises when the influencers manage to drag the opinions of the population in their area. For this last case we also compare in Figure~\ref{fig:pop}d the asymptotic distributions of the popularity computed analytically from~\eqref{eq:h.Pareto} and by numerical integration of model~\eqref{eq:system_Boltzmann} up to the computational time $T=700$. Remarkably function~\eqref{eq:h.Pareto}, in spite of having been obtained under the simplification of steady opinion dynamics, cf. Section~\ref{sect:popularity}, turns out to be a good approximation of the actual popularity distribution computed from model~\eqref{eq:system_Boltzmann} in which popularity trends are dynamically coupled with opinion formation.

\begin{table}[!t]
\caption{Total number of social interactions involving three hashtags in the most popular social networks (Facebook, Twitter, Instagram, Google+, and Youtube) in the period March-April $2018$. Source: \texttt{brand24.com}.}
\label{tab:data}
\begin{center}
\begin{tabular}{|c|ccc|}
\hline
Week & \texttt{\#metoo} & \texttt{\#shareacoke} & \texttt{\#cambridgeanalytica} \\
\hline\hline
$10$ March & $17$ & -- & -- \\
\hline
$17$ March & $10521$ & -- & $468$ \\
\hline
$24$ March & $14954$ & $194$ & $10662$ \\
\hline
$31$ March & $16375$ & $306$ & $9768$ \\
\hline
$7$ April & $15936$ & $594$ & $10055$ \\
\hline
$14$ April & $15337$ & $723$ & $13497$ \\
\hline 
$21$ April & $0$ & $1293$ & $6626$ \\
\hline
\end{tabular}
\end{center}
\end{table}
\begin{figure}[!t]
\subfigure[]{\includegraphics[scale=0.5]{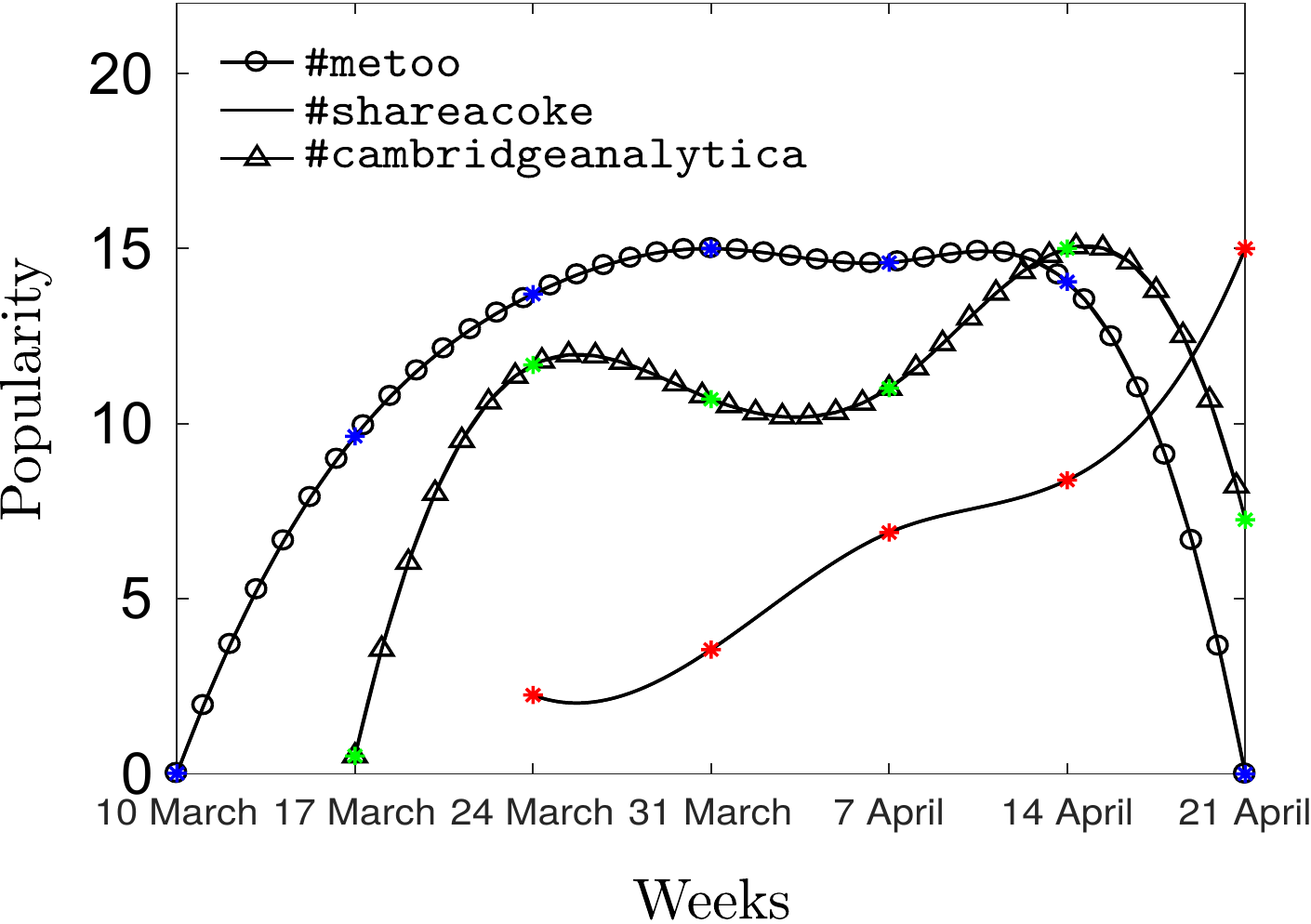}}
\subfigure[]{\includegraphics[scale=0.5]{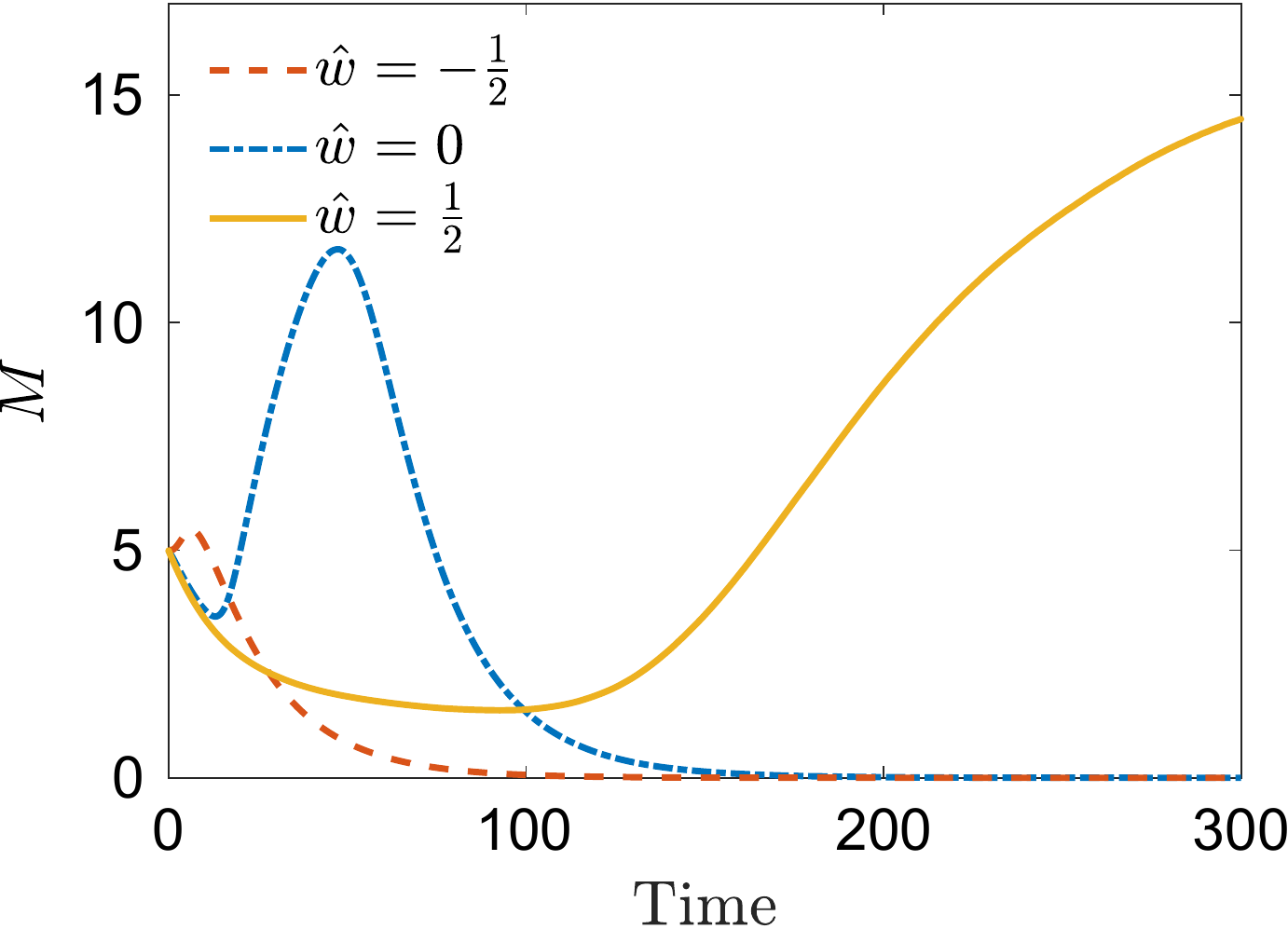}}
\caption{\textbf{Test 3, Section~\ref{sect:test3}}. (a) Third order spline interpolation of the popularity time trend of the three hashtags reported in Table~\ref{tab:data}. For ease of representation data have been normalized with respect to their maxima. Colored markers are the non-interpolated data. (b) Time evolution of the mean popularity $M$ in the three scenarios (a)--(c) considered in Figure~\ref{fig:pop}. }
\label{fig:comp}
\end{figure}

Among the possible measures of popularity in social networks, the number of social interactions involving hashtags (such as e.g., posts on online social platforms, their shares and likes) is adopted by several media-marketing websites. As an example, in Table~\ref{tab:data} we report the week data collected by the reference webapp \texttt{brand24.com} in the period March-April $2018$ concerning three hashtags which have characterized recent social media campaigns:
\begin{enumerate}[label=(\roman*)]
\item \texttt{\#metoo} about the movement against sexual harassment;
\item \texttt{\#shareacoke} about the advertisement of a popular soft drink;
\item \texttt{\#cambridgeanalytica} about the improper use of personal data of social network users.
\end{enumerate}
In Figure~\ref{fig:comp}a we plot the time trends of the popularity of these hashtags obtained by means of a third order spline interpolation of the data in Table~\ref{tab:data}. We point out that data in the plot have been normalized with respect to their maxima for an easier visualization. In Figure~\ref{fig:comp}b we show instead the time evolution of the mean popularity in the scenarios (a)--(c) illustrated in Figure~\ref{fig:pop}. The comparison shows that the popularity trends simulated by our model capture qualitatively well the observable characteristics of representative real trends, thereby confirming that our model may explain the basic microscopic mechanisms of popularity spreading in connection with opinion sharing on social networks.

\section{Conclusions}
In this paper we introduced mathematical models of kinetic type able to follow the marketing of products by using social networks. Our analysis is twofold. First, we considered the problem of opinion formation in a social network in which the change of opinion also depends on the connectivity of the agents. Specifically, in agreement with the literature, we assumed that agents with low connectivity are more prone to be influenced by well connected agents (the so-called influencers). Then, we investigated the spreading of the popularity of a selected product (such as e.g., news, video, advertisement) by coupling it to the opinion dynamics taking place over the social network. In particular, we assumed that the product launched onto the social network targets a given opinion and can be possibly reposted by the agents that it reaches depending on how much their current opinion is aligned with the targeted one. By means of analytical and numerical results we were able to recover the typical fat-tailed statistical distribution of the popularity well acknowledged by empirical observations. In addition to this, our results pointed out the importance of the agents with high connectivity, and of their opinions, in the possible success of the product under consideration. More in general, they highlighted clearly the role of the social network in promoting different forms of consensus, and consequently different trends of popularity, depending on the connectivity distribution of the agents. Finally, we showed the potential effectiveness of our modeling approach by means of a qualitative comparison of the simulated popularity trends with real trends of a few media campaigns.

Further extensions of the proposed model may include, for instance, a variable target opinion which follows the ongoing opinion dynamics over the social network as well as a dynamic network in which connections among the agents can possibly change in time, see e.g.~\cite{albi2017KRM}. Moreover the notion of popularity, here considered especially in connection with product marketing, may be thought of more in general as a measure of the permeation of a given message into the population of agents. Therefore the approach developed in this paper can constitute a formal basis for the investigation of effective communication strategies in awareness campaigns concerning important contemporary social issues such as e.g., homeland security, national health or other social programs.

\section*{Acknowledgements}
This work has been written within the activities of GNFM (Gruppo Nazionale per la Fisica Matematica) and GNCS (Gruppo Nazionale per il Calcolo Scientifico) groups of INdAM (National Institute of High Mathematics). G. T. acknowledges support from the MIUR project ``Optimal mass transportation, geometrical and functional inequalities with applications'' and from IMATI institute of the National Council for Research (Pavia, Italy). M. Z. acknowledges support from ``Compagnia di San Paolo'' (Torino, Italy).


\end{document}